\documentclass[submission, Phys]{SciPost}
\usepackage{braket}
\usepackage{listings}
\usepackage{algorithm}
\usepackage{algpseudocode}

\newcommand{\tr}[0]{\operatorname{tr}}

\newcommand{\tgt}[0]{\text{tgt}}

\renewcommand{\Braket}[2]{\left\langle{}#1\vphantom{#2}\mid{}#2\vphantom{#1}\right\rangle}

\newcommand{\Op}[1]{\hat{#1}}
\newcommand{\dd}[0]{\,\text{d}}
\newcommand{\Liouville}[0]{\mathcal{L}}

\newcommand{\Norm}[1]{\left\lVert#1\right\rVert}
\newcommand{\Abs}[1]{\left\vert#1\right\vert}

\renewcommand{\Re}[0]{\operatorname{Re}}
\renewcommand{\Im}[0]{\operatorname{Im}}

\definecolor{yaleblue}{rgb}{0.06, 0.3, 0.57}
\definecolor{tropicalrainforest}{rgb}{0.0, 0.46, 0.37}

\algrenewcommand\algorithmicrequire{\textbf{Input:}}
\algrenewcommand\algorithmicensure{\textbf{Output:}}
\algnewcommand{\IIf}[1]{\State\algorithmicif\ #1\ \algorithmicthen}
\floatname{algorithm}{Algorithm}
\definecolor{clVarScalar}{rgb}{0.06, 0.3, 0.57} 
\definecolor{clVarState}{rgb}{0.57, 0.06, 0.43} 
\definecolor{clVarOperator}{rgb}{0.09, 0.4, 0.04} 
\definecolor{clVarArray}{rgb}{0.67, 0.06, 0.11} 
\newcommand{\VarScalar}[1]{{\color{clVarScalar}#1}}
\newcommand{\VarState}[1]{\ensuremath{\textcolor{clVarState}{#1}}}
\newcommand{\PropAnnotation}[1]{\textcolor{black!30}{#1}}
\newcommand{\VarPropState}[4]{\ensuremath{\VarState{#1}_{\textcolor{clVarState}{#2}}^{\PropAnnotation{#3}}\PropAnnotation{(#4)}}}  
\newcommand{\VarOperator}[2]{{\color{clVarOperator}#1_{#2}}} 
\newcommand{\VarArray}[1]{{\color{clVarArray}#1}}
\newcommand{\Forall}{\boldsymbol{\forall}}

\hypersetup{%
   pdftitle={Krotov: A Python implementation of Krotov's method for quantum
   optimal control},
   pdfsubject={Krotov - Python},
   pdfauthor={Michael Goerz, Christiane Koch},
   pdfstartpage=1,
   pdfpagemode=UseNone,
   pdfstartview=Fit,
   colorlinks=true,
   linkcolor=black,
   citecolor=blue,
   filecolor=black,
   urlcolor=black
 }

\begin{document}

\begin{center}{\Large \textbf{%
    Krotov:
    A Python implementation of \\Krotov's method for quantum optimal control
}}\end{center}

\begin{center}
M. H. Goerz\textsuperscript{1*},
D. Basilewitsch\textsuperscript{2},
F. Gago-Encinas\textsuperscript{2},
M. G. Krauss\textsuperscript{2},
K. P. Horn\textsuperscript{2},
D. M. Reich\textsuperscript{2},
C. P. Koch\textsuperscript{2,3}
\end{center}

\begin{center}
{\bf 1} U.S. Army Research Lab, Computational and Information Science Directorate, Adelphi, MD 20783, USA
\\
{\bf 2} Theoretische Physik, Universität Kassel, Heinrich-Plett-Str. 40, D-34132 Kassel, Germany
\\
{\bf 3} Dahlem Center for Complex Quantum Systems and Fachbereich Physik, Freie Universität Berlin, Arnimallee 14, 14195 Berlin, Germany
\\~\\
* mail@michaelgoerz.net
\end{center}

\begin{center}
\today
\end{center}


\section*{Abstract}
{\bf
We present a new open-source Python package, \emph{krotov}, implementing
the quantum optimal control method of that name.
It allows to determine time-dependent external fields for a wide range of
quantum control problems, including state-to-state transfer, quantum gate
implementation and optimization towards an arbitrary perfect entangler.
Krotov's method compares to other gradient-based optimization methods
such as gradient-ascent and guarantees monotonic convergence for
approximately time-continuous control fields.
The user-friendly interface allows for combination with other Python packages,
and thus high-level customization.
}

\vspace{10pt}
\noindent\rule{\textwidth}{1pt}
\tableofcontents\thispagestyle{fancy}
\noindent\rule{\textwidth}{1pt}
\vspace{10pt}

\section{Introduction}%
\label{sec:intro}

Quantum information science has changed our perception of quantum physics from
passive understanding to a source of technological advances~\cite{AcinNJP18}.
By way of actively exploiting the two essential elements of quantum physics,
coherence and entanglement, technologies such as quantum
computing~\cite{NielsenChuang} or quantum sensing~\cite{DegenRMP17} hold the
promise for solving computationally hard problems or reaching unprecedented
sensitivity.
These technologies rely on the ability to accurately perform quantum operations
for increasingly complex quantum systems.
Quantum optimal control allows to address this challenge by providing a set of
tools to devise and implement shapes of external fields that accomplish a given
task in the best way possible~\cite{GlaserEPJD2015}.
Originally developed in the context of molecular
physics~\cite{Tannor92,GrossJCP92} and nuclear magnetic
resonance~\cite{MurdochJMR87,GlaserCPL89}, quantum optimal control theory has
been adapted to the specific needs of quantum information science in recent
years~\cite{GlaserEPJD2015,KochJPCM16}.
Calculation of optimized external field shapes for tasks such as state
preparation or quantum gate implementation have thus become
standard~\cite{GlaserEPJD2015}, even for large Hilbert space dimensions as
encountered in e.g.\ Rydberg atoms~\cite{CuiQST17,PatschPRA18}.
Experimental implementation of the calculated field shapes, using arbitrary
waveform generators, has been eased by the latter becoming available
commercially.
Successful demonstration of quantum operations in various
experiments~\cite{GlaserEPJD2015,LovecchioPRA16,vanFrankSciRep16,OfekNat16,SorensenNat16,HeeresNatComm17,HeckPNAS18,FengPRA18,OmranS2019,Larrouy}
attests to the level of maturity that quantum optimal control in quantum
technologies has reached.

In order to calculate optimized external field shapes, two choices need to be
made -- about the optimization functional and about the optimization method.
The functional consists of the desired figure of merit, such as a gate or state
preparation error, as well as additional constraints, such as amplitude or
bandwidth restrictions~\cite{GlaserEPJD2015,KochJPCM16}.
Optimal control methods in general can be classified into gradient-free and
gradient-based algorithms that either evaluate the optimization functional alone
or together with its gradient~\cite{GlaserEPJD2015}.
Gradient-based methods typically converge faster, unless the number of
optimization parameters can be kept small.
Most gradient-based methods rely on the iterative solution of a set of coupled
equations that include forward propagation of initial states, backward
propagation of adjoint states, and the control update~\cite{GlaserEPJD2015}.
A popular representative of concurrent update methods is GRadient Ascent Pulse
Engineering (GRAPE)~\cite{KhanejaJMR05}.
Krotov's method, in contrast, requires sequential
updates~\cite{Tannor92,ReichJCP12}.
This comes with the advantage of guaranteed monotonic convergence and obviates
the need for a line search in the direction of the gradient~\cite{EitanPRA11}.
While GRAPE is found in various software packages, there has not been an open
source implementation of Krotov's method to date.
Our package provides that missing implementation.

The choice of Python as an implementation language is due to Python's
easy-to-learn syntax, expressiveness, and immense popularity in the scientific
community.
Moreover, the QuTiP library~\cite{JohanssonCPC2012, JohanssonCPC2013}  exists,
providing a general purpose tool to numerically describe quantum systems and
their dynamics.
QuTiP already includes basic versions of other popular quantum control
algorithms such as GRAPE and the gradient-free CRAB~\cite{CanevaPRA11}.
The Jupyter notebook framework~\cite{Jupyter} is available to provide an ideal
platform for the interactive exploration of the \texttt{krotov} package's
capabilities, and to facilitate reproducible research workflows.

The \texttt{krotov} package presented herein targets both students wishing to
enter the field of quantum optimal control, and researchers in the field.
By providing a comprehensive set of examples, we enable users of our package to
explore the formulation of typical control problems, and to understand how
Krotov's method can solve them.
These examples are inspired by recent
publications~\cite{MullerQIP11,GoerzPRA2014,GoerzNJP2014,WattsPRA2015,
GoerzPRA2015,BasilewitschNJP2017}, and thus show the use of the method in the
purview of current research.
In particular, the package is not restricted to closed quantum systems, but can
fully address open system dynamics, and thus aide in the development of Noisy
Intermediate-Scale Quantum (NISQ) technology~\cite{PreskillQ2018}.
Optimal control is also increasingly important in the design of
experiments~\cite{GlaserEPJD2015,LovecchioPRA16,vanFrankSciRep16,OfekNat16,SorensenNat16,HeeresNatComm17,HeckPNAS18,FengPRA18,OmranS2019,Larrouy},
and we hope that the availability of an easy-to-use implementation of Krotov's
method will facilitate this further.

Large Hilbert space
dimensions~\cite{GoerzEPJQT2015,GoerzNPJQI17,CuiQST17,PatschPRA18} and open
quantum systems~\cite{GoerzNJP2014} in particular require considerable numerical
effort to optimize.
Compared to the Fortran and C/C++ languages traditionally used for scientific
computing, and more recently Julia~\cite{BezansonSIREV2017}, pure Python code
usually performs slower by two to three orders of
magnitude~\cite{AkeretAC2015,EichhornCSJ2018}.
Thus, for hard optimization problems that require several thousand iterations to
converge, the Python implementation provided by the \texttt{krotov} package may
not be sufficiently fast.
In this case, it may be desirable to implement the entire optimization and time
propagation in a single, more efficient (compiled) language.
Our Python implementation of Krotov's method puts an emphasis on clarity, and
the documentation provides detailed explanations of all necessary concepts,
especially the correct time discretization, see
Appendix~\ref{apx:krotov_update_discretization}, and the possibility to
parallelize the optimization.
Thus, the \texttt{krotov} package can serve as a reference implementation,
leveraging Python's reputation as ``executable pseudocode'', and as a foundation
against which to test other implementations.

This paper is structured as follows: In Sec.~\ref{sec:overview}, we give a brief
overview of Krotov's method as it is implemented in the package.
Based on a simple example, the optimization of a state-to-state transition in a
two-level system, we describe the interface of the \texttt{krotov} package and
its capabilities.
Section~\ref{sec:common_optimization_tasks} goes beyond that simple example to
discuss how the \texttt{krotov} package can be used to solve some common, more
advanced problems in quantum optimal control, involving complex-valued control
fields, optimization of quantum gates in Hilbert or Liouville space,
optimization over an ensemble of noise realizations, and  use of non-convex
functionals which occur e.g.\ in the optimization towards an arbitrary perfect
entangler.
Section~\ref{sec:comparison} compares Krotov's method to other methods commonly
used in quantum optimal control, in order to provide guidance on when use of the
\texttt{krotov} package is most appropriate.
Section~\ref{sec:future_perspectives} presents future perspectives, and
Section~\ref{sec:conclusions} concludes.
Appendix~\ref{apx:krotov_update} defines and explains the time-discretized
update equation that underlies the implementation of Krotov's method.
Appendix~\ref{apx:pseudocode} gives a detailed technical specification of the
optimization algorithm in pseudocode format, and analyzes the required numerical
resources with respect to CPU time and memory.
Appendices~\ref{apx:installation_instructions}~and~\ref{apx:package_docs}
contain installation instructions for the \texttt{krotov} package and link to
its online documentation.

\section{Overview of Krotov's method and the \texttt{krotov} package}%
\label{sec:overview}

\subsection{The quantum control problem}%
\label{sec:overview_control_problem}

Quantum optimal control methods formalize the problem of finding ``control
fields'' that steer the time evolution of a quantum system in some desired way.
For closed systems, described by a Hilbert space state $\ket{\Psi(t)}$, this
time evolution is given by the Schrödinger equation,
\begin{equation}
  \frac{\partial}{\partial t} \Ket{\Psi(t)}
  = -\frac{\mathrm{i}}{\hbar} \Op{H}(t)\Ket{\Psi(t)}\,,
\end{equation}
where the Hamiltonian $\Op{H}(t)$ depends on one or more control fields
$\{\epsilon_l(t)\}$.
We often assume the Hamiltonian to be linear in the controls,
\begin{equation}
  \Op{H}(t)
  = \Op{H}_0 + \epsilon_1(t) \Op{H}_1 + \epsilon_2(t) \Op{H}_2 + \dots
\end{equation}
but non-linear couplings may also occur, for example when considering
non-resonant multi-photon transitions.
For open quantum systems described by a density matrix $\hat{\rho}(t)$, the
Liouville-von-Neumann equation
\begin{equation}
  \frac{\partial}{\partial t} \hat{\rho}(t) = \frac{1}{\hbar} \Liouville(t) \hat{\rho}(t)
\end{equation}
replaces the Schrödinger equation, with the (non-Hermitian) Liouvillian
$\Liouville(t)$.
The most direct example of a control problem is a state-to-state transition.
The objective is for a known quantum state $\ket{\phi}$ at time zero to evolve
to a specific target state $\ket{\phi^\tgt}$ at final time $T$, controlling,
e.g.\ a chemical reaction~\cite{TannorJCP1985}.
Another example is the realization of quantum gates, the building blocks of a
quantum computer.
In this case, the states forming a computational basis must transform according
to a unitary transformation~\cite{NielsenChuang}, see
Section~\ref{sec:task_gate_oct}.
Thus, the control problem involves not just the time evolution of a single
state, but a set of states $\{\ket{\phi_k(t)}\}$.
Generalizing even further, each state $\ket{\phi_k(t)}$ in the control problem
may evolve under a different Hamiltonian $\Op{H}_k(\{\epsilon_l(t)\})$, see
Section~\ref{sec:task_ensemble_oct}.

Physically, the control fields $\{\epsilon_l(t)\}$ might be the amplitudes of a
laser pulse for the control of molecular systems or trapped atom/ion quantum
computers, radio-frequency fields for nuclear magnetic resonance, or microwave
fields for superconducting circuits.
When there are multiple independent controls $\{\epsilon_l(t)\}$ involved in the
dynamics, these may correspond e.g., to different color lasers used in the
excitation of a Rydberg atom, or different polarization components of an
electric field.

The quantum control methods build on a rich field of classical control
theory~\cite{BellmanBook,PontryaginBook}.
This includes Krotov's
method~\cite{KrotovEC1983,KrotovCC1988,Krotov.book,KonnovARC99}, which was
originally formulated to optimize the soft landing of a spacecraft from orbit to
the surface of a planet, before being applied to quantum mechanical
problems~\cite{Tannor92,SomloiCP1993,BartanaJCP1997,SklarzPRA2002,ReichJCP12}.
Fundamentally, they rely on the variational principle, that is, the minimization
of a functional $J[\{\ket{\phi_k^{(i)}(t)}\}, \{\epsilon_l^{(i)}(t)\}]$ that
includes any required constraints via Lagrange multipliers.
The condition for minimizing $J$ is then $\nabla_{\phi_k, \epsilon_l} J = 0$.
In rare cases, the variational calculus can be solved in closed form, based on
Pontryagin's maximum principle~\cite{PontryaginBook}.
Numerical methods are required in any other case.
These start from an initial guess control (or set of guess controls, if there
are multiple controls), and calculate an update to these controls that will
decrease the value of the functional.
The updated controls then become the guess for the next iteration of the
algorithm, until the value of the functional is sufficiently small, or
convergence is reached.

\subsection{Optimization functional}%
\label{sec:overview_functional}

Mathematically, Krotov's method, when applied to quantum
systems~\cite{Tannor92,ReichJCP12}, minimizes a functional of the most general
form
\begin{equation}
  \label{eq:functional}
  J[\{\ket{\phi_k^{(i)}(t)}\}, \{\epsilon_l^{(i)}(t)\}]
    = J_T(\{\ket{\phi_k^{(i)}(T)}\})
        + \sum_l \int_0^T g_a(\epsilon_l^{(i)}(t)) \dd t
        + \int_0^T g_b(\{\phi^{(i)}_k(t)\}) \dd t\,,
\end{equation}
where the \(\{\ket{\phi_k^{(i)}(T)}\}\) are the time-evolved initial states
\(\{\ket{\phi_k}\}\) under the controls \(\{\epsilon^{(i)}_l(t)\}\) of the
$i$'th iteration.
In the simplest case of a single state-to-state transition, the index $k$
vanishes.
For the example of a two-qubit quantum gate, \(\{\ket{\phi_k}\}\) would be the
logical basis states $\ket{00}$, $\ket{01}$, $\ket{10}$, and $\ket{11}$, all
evolving under the same Hamiltonian $\Op{H}_k \equiv \Op{H}$.
The sum over $l$ vanishes if there is only a single control.
For open system dynamics, the states \(\{\ket{\phi_k}\}\) may be density
matrices.

The functional consists of three parts:
\begin{itemize}
\item
  A final time functional \(J_T\).
  This is the ``main'' part of the functional, and we can usually think of \(J\)
  as being an auxiliary functional in the optimization of \(J_T\).
  The most straightforward final time functional for a simple state-to-state
  transition $\ket{\phi} \rightarrow \ket{\phi^{\tgt}}$ is~\cite{PalaoPRA2003}
  \begin{equation}%
    \label{eq:JTss}
    J_{T,\text{ss}} = 1 - \Abs{\Braket{\phi^\tgt}{\phi(T)}}^2\,,
  \end{equation}
  where $\ket{\phi(T)}$ is the time evolution of $\ket{\phi}$ to final time $T$.
  For a quantum gate $\Op{O}$, a typical functional is~\cite{PalaoPRA2003}
  \begin{equation}%
    \label{eq:JTre}
    J_{T,\text{re}} = 1- \frac{1}{N} \Re\left[ \sum_{k=1}^{N} \tau_k \right]\,,
    \quad\text{with}\quad
    \tau_k = \Braket{\phi_k^\tgt}{\phi_k(T)}\,,\quad
    \ket{\phi_k^{\tgt}} = \Op{O} \ket{\phi_k}\,,
  \end{equation}
  and $N$ being the dimension of the logical subspace, e.g. $N=4$
  and $\{\ket{\phi_k}\} = \{\ket{00}, \ket{01}$, $\ket{10}, \ket{11} \}$ for
  a two-qubit gate.
  The use of the real part in the functional implies that we
  care about the global phase of the achieved gate.
\item
  A running cost on the control fields, \(g_a\).
  The most commonly used expression (and the only one currently supported by the
  \texttt{krotov} package) is~\cite{PalaoPRA2003}
  \begin{equation}
  \begin{split}
    \label{eq:g_a}
    g_a(\epsilon_l^{(i)}(t))
    &= \frac{\lambda_{a,l}}{S_l(t)} \left(
        \epsilon_l^{(i)}(t) - \epsilon_{l, \text{ref}}^{(i)}(t)
      \right)^2\,;
      \quad
    \epsilon^{(i)}_{l, \text{ref}}(t) = \epsilon_l^{(i-1)}(t)\\
   &= \frac{\lambda_{a,l}}{S_l(t)} \left( \Delta\epsilon_l^{(i)}(t) \right)^2
   \,,
  \end{split}
  \end{equation}
  with the inverse ``step width'' $\lambda_{a,l} > 0$, the ``update shape''
  function $S_{l}(t) \in [0, 1]$, and the control update
  \begin{equation}%
    \label{eq:update}
    \Delta\epsilon_l^{(i)}(t)
    \equiv \epsilon_l^{(i)}(t) - \epsilon_l^{(i-1)}(t)\,,
  \end{equation}
  where $\epsilon_l^{(i-1)}(t)$ is the optimized control of the previous
  iteration -- that is, the guess control of the current iteration $(i)$.
\item
  An optional state-dependent running cost, \(g_b\).
  This may be used to encode time-dependent control targets~\cite{KaiserJCP2004,
  SerbanPRA2005}, or to penalize population in a subspace~\cite{PalaoPRA2008}.
  The presence of a state-dependent constraint in the functional entails an
  inhomogeneous term in the backward propagation in the calculation of the
  control updates in each iteration of Krotov's method, see
  Appendix~\ref{apx:krotov_update}, and is currently not supported by the
  \texttt{krotov} package.
  Penalizing population in a subspace can also be achieved through simpler
  methods that do not require a $g_b$, e.g., by using
  a non-Hermitian Hamiltonian to remove population from the forbidden subspace
  during the time evolution.
\end{itemize}

\subsection{Iterative control update}%
\label{sec:overview_update}

Starting from the initial guess control $\epsilon_l^{(0)}(t)$, the optimized
field $\epsilon_l^{(i)}(t)$ in iteration $i > 0$ is the result of applying
a control update,
\begin{equation}%
  \label{eq:eps_updated}
  \epsilon_l^{(i)}(t)
  = \epsilon_l^{(i-1)}(t) + \Delta\epsilon_l^{(i)}(t)\,.
\end{equation}
Krotov's method is a clever construction of a particular
$\Delta\epsilon_l^{(i)}(t)$ that ensures
\begin{equation*}
  J[\{\ket{\phi_k^{(i)}(t)}\}, \{\epsilon_l^{(i)}(t)\}] \leq
  J[\{\ket{\phi_k^{(i-1)}(t)}\}, \{\epsilon_l^{(i-1)}(t)\}]\,.
\end{equation*}
Krotov's solution for $\Delta\epsilon_l^{(i)}(t)$ is given
in Appendix~\ref{apx:krotov_update}.
As shown there, for the specific running cost of Eq.~\eqref{eq:g_a}, using the
guess control field $\epsilon_l^{(i-1)}(t)$ as the ``reference'' field, the
update $\Delta\epsilon^{(i)}_l(t)$ is proportional to
$\frac{S_l(t)}{\lambda_{a,l}}$.
Note that this also makes $g_a$ proportional to
$\frac{S_l(t)}{\lambda_{a,l}}$, so that Eq.~\eqref{eq:g_a} is still well-defined
for $S_l(t) = 0$.
The (inverse) Krotov step width \(\lambda_{a,l}\) can be used to
determine the overall magnitude of $\Delta\epsilon^{(i)}_l(t)$.
Values that are too large will change \(\epsilon_l^{(i)}(t)\) by only a small
amount in every iteration, causing slow convergence.
Values that are too small will result in numerical instability, see
Appendix.~\ref{apx:krotov_update_discretization}.
The update shape function $S_l(t)$ allows to ensure boundary
conditions on $\epsilon^{(i)}_l(t)$: If both the guess field
$\epsilon^{(i-1)}_l(t)$ and $S_l(t)$ switch on and off smoothly around $t=0$
and $t=T$, then this feature will be preserved by the optimization.
A typical example for an update shape is
\begin{equation}%
  \label{eq:flattop}
  S_l(t) = \begin{cases}
    B(t; t_0=0, t_1=2 t_{\text{on}})
      & \text{for} \quad 0 < t < t_{\text{on}} \\
    1 & \text{for} \quad t_{\text{on}} \le t \le T - t_{\text{off}} \\
    B(t; t_0=T-2 t_{\text{off}}, t_1=T)
      & \text{for} \quad T - t_{\text{off}} < t < T\,,
  \end{cases}
\end{equation}
with the Blackman shape
\begin{equation}%
  \label{eq:blackman}
  B(t; t_0, t_1) =
    \frac{1}{2}\left(
      1 - a - \cos\left(2\pi \frac{t - t_0}{t_1 - t_0}\right)
      + a \cos\left(4\pi \frac{t - t_0}{t_1 - t_0}\right)
    \right)\,,\quad a = 0.16\,,
\end{equation}
which is similar to a Gaussian, but exactly zero at $t = t_0, t_1$.
This is essential to maintain the typical boundary condition of zero
amplitude at the beginning and end of the optimized control field.
Generally, \emph{any} part of the control field can be kept unchanged in the
optimization by choosing $S_l(t) = 0$ for the corresponding intervals of the
time grid.

\subsection{Example: state-to-state transition in a two-level system}%
\label{sec:overview_example}

As a first taste of the \texttt{krotov} package's usage, we consider a simple
but complete example for the optimization of a state-to-state optimization in
Hilbert space, specifically the transformation $\ket{0} \rightarrow \ket{1}$ in
a two-level system $\Op{H} = -\frac{\omega}{2} \Op{\sigma}_z + \epsilon(t)
\Op{\sigma}_x$, where $\Op{\sigma}_z$ and $\Op{\sigma}_x$ are the Pauli-z and
Pauli-x matrices, respectively, $\omega$ is the transition frequency between the
levels $\ket{0}$ and $\ket{1}$, and $\epsilon(t)$ is the control field.
In the language of quantum computing, we are attempting to realize a
bit-flip of a qubit from zero to one.
The example assumes that the \texttt{krotov} package and other prerequisites
have been installed on the user's system, see
Appendix~\ref{apx:installation_instructions}.
The full example script, as well as a Jupyter notebook version are also
available as part of the package's online documentation, along with additional
examples, see Appendix~\ref{apx:package_docs}.

\vspace{12pt}
\begin{lstlisting}[language=python,firstline=1,lastline=107,numbers=left]
#!/usr/bin/env python
"""Example script for the optimization of a simple state-to-state
transition in a two-level system"""
import krotov
import qutip
import numpy as np


# First, we define the physical system (a simple TLS)

def hamiltonian(omega=1.0, ampl0=0.2):
    """Two-level-system Hamiltonian

    Args:
        omega (float): energy separation of the qubit levels
        ampl0 (float): constant amplitude of the driving field
    """
    H0 = -0.5 * omega * qutip.operators.sigmaz()
    H1 = qutip.operators.sigmax()

    def guess_control(t, args):
        return ampl0 * krotov.shapes.flattop(
            t, t_start=0, t_stop=5, t_rise=0.3, func="blackman"
        )

    return [H0, [H1, guess_control]]


H = hamiltonian()
tlist = np.linspace(0, 5, 500)


# Second, we define the control objective: a state-to-state
# transition from the |0⟩ eigenstate to the |1⟩ eigenstate

objectives = [
    krotov.Objective(
        initial_state=qutip.ket("0"), target=qutip.ket("1"), H=H
    )
]


# The magnitude of the pulse updates at each point in time are
# determined by the Krotov step size lambda_a and the
# time-dependent update shape (in [0, 1])
def S(t):
    """Shape function for the field update"""
    return krotov.shapes.flattop(
        t, t_start=0, t_stop=5, t_rise=0.3, func="blackman"
    )


# set required parameters for H[1][1] (the guess_control)
pulse_options = {H[1][1]: dict(lambda_a=5, update_shape=S)}


# Before performing the optimization, it is usually a good idea
# to observe the system dynamics under the guess pulse. The
# mesolve method of the objective delegates to QuTiP's mesolve,
# and can calculate the expectation values of the projectors
# onto the |0⟩ and |1⟩ states, i.e., the population.

proj0, proj1 = (qutip.ket2dm(qutip.ket(l)) for l in ("0", "1"))
e_ops = [proj0, proj1]
guess_dynamics = objectives[0].mesolve(tlist, e_ops=e_ops)

# the resulting expectations values are in guess_dynamics.expect.
# The final-time populations are:

print(
    "guess final time population in |0⟩, |1⟩: %.3f, %.3f\n"
    % tuple([guess_dynamics.expect[l][-1] for l in (0, 1)])
)


# Now, we perform the actual optimization

opt_result = krotov.optimize_pulses(
    objectives,
    pulse_options=pulse_options,
    tlist=tlist,
    propagator=krotov.propagators.expm,
    chi_constructor=krotov.functionals.chis_ss,
    info_hook=krotov.info_hooks.print_table(
        J_T=krotov.functionals.J_T_ss
    ),
    check_convergence=krotov.convergence.Or(
        krotov.convergence.value_below('1e-3', name='J_T'),
        krotov.convergence.check_monotonic_error,
    ),
    store_all_pulses=True,
)

print("\n", opt_result, sep='')


# We can observe the population dynamics under the optimized
# control

opt_dynamics = opt_result.optimized_objectives[0].mesolve(
    tlist, e_ops=[proj0, proj1]
)

print(
    "\noptimized final time population in |0⟩, |1⟩: %.3f, %.3f"
    % (opt_dynamics.expect[0][-1], opt_dynamics.expect[1][-1])
)
\end{lstlisting}

The example starts by importing the \texttt{krotov} package, as well as QuTiP
(the ``Quantum Toolbox in Python'')~\cite{JohanssonCPC2012, JohanssonCPC2013}
and NumPy (the standard package providing numeric arrays in
Python)~\cite{van-der-WaltCSE2011}, used here to specify the propagation time
grid.
The integration of the \texttt{krotov} package with QuTiP is central: All
operators and states are expressed as \texttt{qutip.Qobj} objects.
Moreover, the \texttt{optimize\_pulses} interface for Krotov's optimization
method is inspired by the interface of QuTiP's central \texttt{mesolve} routine
for simulating the system dynamics of a closed or open quantum system.
In particular, when setting up an optimization, the (time-dependent) system
Hamiltonian should be represented by a nested list.
That is, a Hamiltonian of the form \(\Op{H} = \Op{H}_0 + \epsilon(t) \Op{H}_1\)
is represented as \texttt{H = [H0, [H1, eps]]} where \texttt{H0} and \texttt{H1}
are \texttt{qutip.Qobj} operators, and \texttt{eps} representing \(\epsilon(t)\)
is a function with signature \texttt{eps(t,\ args)}, or an array of control
values with the length of the time grid (\texttt{tlist} parameter). The
\texttt{hamiltonian} function in line~11 of the example sets up exactly such an
operator, using a control field with a flattop/Blackman envelope as specified in
Eqs.~(\ref{eq:flattop},~\ref{eq:blackman}).

The next steps in the example set up the arguments required for the optimization
initiated in line~78.
The \texttt{optimize\_pulses} function is the central routine provided by the
\texttt{krotov} package.
Its most important parameters are
\begin{itemize}
    \item \texttt{objectives}: a list of objectives, each of which is an
      instance of \texttt{krotov.Objective}. Each objective has an
      \texttt{initial\_state}, which is a \texttt{qutip.Qobj} representing
      a Hilbert space state or density matrix, a \texttt{target} (usually the
      target state that the initial state should evolve into when the objective
      is fulfilled), and a Hamiltonian or Liouvillian \texttt{H} in the
      nested-list format described above. In this example, there is a single
      objective for the transition $\ket{0} \rightarrow \ket{1}$ under the
      Hamiltonian initialized in line~29. The objectives express the goal of the
      optimization \emph{physically}. However, they do not fully specify the
      functional $J_T$ that encodes the goal of the optimization
      \emph{mathematically}: instead, $J_T$ is implicit in the
      \texttt{chi\_constructor} argument, see below.
    \item \texttt{pulse\_options}: a dictionary that maps each control to the
      parameters $\lambda_{a,l}$ (the Krotov update step size) and $S_l(t)$ (the
      update shape). In this example, \texttt{H[1][1]} refers to the
      \texttt{guess\_control} in line~21. The value of 5 for $\lambda_a$
      (no index $l$, as there is only a single control) was chosen by trial
      and error. $S(t)$ corresponds to the function defined in
      Eqs.~(\ref{eq:flattop},~\ref{eq:blackman}). The fact that $S(t)$ is the
      same formula as the envelope of the \texttt{guess\_control} is
      incidental: $S(t)$ as the \texttt{update\_shape} in the
      \texttt{pulse\_options} only scales the \emph{update} of the control field
      in each iteration, in this case enforcing that the value of the optimized
      fields remains zero at initial and final time.
    \item \texttt{tlist}: An array of time grid values in $[0, T]$. Internally,
      the controls are discretized as piecewise-constant on the intervals of
      this time grid. Here, the time grid is initialized in line~30, with 500
      points between $t_0 = 0$ and $T = 5$. This is chosen such that the
      piecewise-constant approximation is sufficiently good to not affect the
      results within the shown precision of three significant digits.
    \item \texttt{propagator}: A routine that calculates the time evolution for
      a state over a single interval of the time grid. This allows the
      optimization to use arbitrary equations of motion. Also,
      since the main numerical effort in the optimization is the forward- and
      backward propagation of the states, the ability to supply a highly
      optimized propagator is key to numerical efficiency. In this example, we
      use the \texttt{expm} propagator that is included in the \texttt{krotov}
      package. It evaluates the result of the time propagation $\ket{\Psi(t+\dd
      t)} = \Op{U} \ket{\Psi(t)}$ by explicitly constructing the time evolution
      operator $\Op{U} = \exp[\mathrm{i} \Op{H} \dd t]$ through
      matrix-exponentiation ($\hbar = 1$). Full matrix-exponentiation is
      inefficient for larger Hilbert space dimensions. For a dimension $>10$ the
      \texttt{expm} propagator can still be useful as an ``exact'' propagator
      for debugging purposes.
    \item \texttt{chi\_constructor}: a function that calculates a set of states
      $\{\ket{\chi_k^{(i-1)}(T)}\}$, according to the equation
      \begin{equation}%
        \label{eq:chi_boundary}
        \Ket{\chi_k^{(i-1)}(T)}
        = - \left.\frac{\partial J_T}{\partial \Bra{\phi_k(T)}}
          \right\vert_{(i-1)}\,,
      \end{equation}
      where the right-hand-side is evaluated for the set of states
      $\{\ket{\phi_k^{(i-1)}(T)}\}$ resulting from the forward-propagation of
      the initial states of the \texttt{objectives} under the guess controls of
      iteration $(i)$ -- that is, the optimized controls of the previous
      iteration $(i-1)$.
      The constructed states $\{\ket{\chi_{k}^{(i-1)}(T)}\}$ then serve as the
      boundary condition for the backward propagation in Krotov's method, see
      Appendices~\ref{apx:krotov_update},~\ref{apx:pseudocode}.
      The \texttt{chi\_constructor} implicitly defines the functional $J_T$: For
      every choice of functional, there is a corresponding
      \texttt{chi\_constructor} that must be implemented from the analytic
      solution of Eq.~\eqref{eq:chi_boundary}.
      The \texttt{krotov} package includes the
      \texttt{chi\_constructor} functions for the most common functionals in
      quantum optimal control.
      Here, \texttt{chis\_ss} matches the functional $J_{T,\text{ss}}$ in
      Eq.~\eqref{eq:JTss},
      \begin{equation}
      \begin{split}
        \ket{\chi^{(i-1)}(T)}
         &= \frac{\partial}{\partial \Bra{\phi(T)}}
             \underbrace{%
               \braket{\phi(T)|\phi^\tgt}
               \braket{\phi^\tgt|\phi(T)}
             }_{\Abs{\Braket{\phi^\tgt}{\phi(T)}}^2}
             \Bigg\vert_{(i-1)} \\
         &= \left(\braket{\phi^\tgt|\phi^{(i-1)}(T)}\right) \Ket{\phi^\tgt}\,.
      \end{split}
      \end{equation}
\end{itemize}

The call to \texttt{optimize\_pulses} also includes two optional arguments that
are used for convergence analysis.
Without these, the optimization would run silently for a fixed number of
iterations and then return a \texttt{Result} object (\texttt{opt\_result} in the
example) that contains the optimized controls discretized to the points of
\texttt{tlist}, alongside other diagnostic data.
The two parameters that allow to keep track of the optimization progress and to
stop the optimization based on this progress, are
\begin{itemize}
\item \texttt{info\_hook}: A function that receives the data resulting from an
iteration of the algorithm, and may print arbitrary diagnostic information and
calculate the value of the functional $J_T$.
Any value returned from the \texttt{info\_hook} will be available in the
\texttt{info\_vals} attribute of the final \texttt{Result} object.
Here, we use an \texttt{info\_hook} that prints a tabular overview of the
functional values and the change in the functional in each iteration, see the
script output below.
This is the only place where $J_T$ is calculated \emph{explicitly}, via the
\texttt{J\_T\_ss} function that evaluates Eq.~\eqref{eq:JTss}.
\item \texttt{check\_convergence}: A function that may stop the optimization
based on the previously calculated \texttt{info\_vals}.
The \texttt{krotov} package includes suitable routines for detecting if the
value of $J_T$, or the change $\Delta J_T$ between iterations falls below a
specified limit.
In the example, we chain two function via \texttt{Or}:
The first function, \texttt{value\_below}, stops the optimization when the value
of $J_{T,\text{ss}}$ falls below $10^{-3}$, and the second function,
\texttt{check\_monotonic\_error}, is a safety check to verify that the value of
$J_{T,\text{ss}}$ decreases in each iteration.
Both of these rely on the value of $J_{T,\text{ss}}$ having been calculated in
the previous \texttt{info\_hook}.
\end{itemize}
The parameter \texttt{store\_all\_pulses} is set here to ensure that the
optimized controls from each iteration will be available in the
\texttt{all\_pulses} attribute of the \texttt{Result},
allowing for a detailed analysis of each iteration after the optimization ends,
cf.~Fig~\ref{fig:tls_example} below.
Without this parameter, only the final optimized controls are kept.
See the Jupyter notebook version of the example
(Appendix~\ref{apx:package_docs}) for details on how to obtain
Fig~\ref{fig:tls_example}.

Before and after the optimization, the \texttt{mesolve} method of the
\texttt{Objective} is used in the example to simulate the dynamics of the system
under the guess control and the optimized control, respectively.
This method delegates directly to QuTiP's \texttt{mesolve} function.

Overall, the example illustrates the general procedure for optimizing with the
\texttt{krotov} package:
\begin{enumerate}
  \item
    define the necessary quantum operators and states using QuTiP objects,
  \item
    create a list of optimization objectives, as instances of
    \texttt{krotov.Objective}, \item call \texttt{krotov.optimize\_pulses} to
    perform an optimization of an arbitrary number of control fields over all
    the objectives.
\end{enumerate}

\vspace{12pt}
{%
\parindent0pt Running the example script generates the following output: }
\vspace{12pt} 
\begin{lstlisting}
guess final time population in |0⟩, |1⟩: 0.951, 0.049

iter.      J_T    ∫gₐ(t)dt          J       ΔJ_T         ΔJ  secs
0     9.51e-01    0.00e+00   9.51e-01        n/a        n/a     1
1     9.24e-01    2.32e-03   9.27e-01  -2.70e-02  -2.47e-02     2
2     8.83e-01    3.53e-03   8.87e-01  -4.11e-02  -3.75e-02     2
3     8.23e-01    5.22e-03   8.28e-01  -6.06e-02  -5.54e-02     2
4     7.38e-01    7.39e-03   7.45e-01  -8.52e-02  -7.78e-02     1
5     6.26e-01    9.75e-03   6.36e-01  -1.11e-01  -1.01e-01     1
6     4.96e-01    1.16e-02   5.07e-01  -1.31e-01  -1.19e-01     1
7     3.62e-01    1.21e-02   3.74e-01  -1.34e-01  -1.22e-01     1
8     2.44e-01    1.09e-02   2.55e-01  -1.18e-01  -1.07e-01     2
9     1.53e-01    8.43e-03   1.62e-01  -9.03e-02  -8.19e-02     1
10    9.20e-02    5.80e-03   9.78e-02  -6.14e-02  -5.56e-02     1
11    5.35e-02    3.66e-03   5.72e-02  -3.85e-02  -3.48e-02     2
12    3.06e-02    2.19e-03   3.28e-02  -2.29e-02  -2.07e-02     1
13    1.73e-02    1.27e-03   1.86e-02  -1.33e-02  -1.20e-02     2
14    9.79e-03    7.24e-04   1.05e-02  -7.55e-03  -6.82e-03     2
15    5.52e-03    4.10e-04   5.93e-03  -4.27e-03  -3.86e-03     2
16    3.11e-03    2.31e-04   3.35e-03  -2.41e-03  -2.18e-03     2
17    1.76e-03    1.30e-04   1.89e-03  -1.36e-03  -1.23e-03     1
18    9.92e-04    7.36e-05   1.07e-03  -7.65e-04  -6.91e-04     1

Krotov Optimization Result
--------------------------
- Started at 2019-11-23 15:31:52
- Number of objectives: 1
- Number of iterations: 18
- Reason for termination: Reached convergence: J_T < 1e-3
- Ended at 2019-11-23 15:32:30 (0:00:38)

optimized final time population in |0⟩, |1⟩: 0.001, 0.999
\end{lstlisting}
The table that makes up the main part of the output is the result of the
\texttt{print\_table} function that was passed as an \texttt{info\_hook} in
line~84 of the example.
The columns are the iteration number, where iteration 0 is an evaluation of the
guess control; the value of the final time functional $J_T = J_{T,\text{ss}}$,
see Eq.~\eqref{eq:JTss}; the value of the running cost with
$g_a(\epsilon_l^{(i)}(t))$ given by Eq.~\eqref{eq:g_a}, which is a measure of
how much the controls change in each iteration and thus allows to gauge
convergence; the value of the total functional $J$ according to
Eq.~\eqref{eq:functional}; the change in the value of $J_T$ relative to the
previous iteration; the change in the total functional $J$; and finally the
wallclock time in seconds spent on that iteration.
The change in the total functional $\Delta J^{(i)}$ is guaranteed to be negative
(monotonic convergence), up to the effects of time discretization.
Note that
\begin{equation}
  \Delta J^{(i)} =
    \Delta J_T^{(i)} + \sum_l \int_0^T
    \frac{\lambda_{a,l}}{S_l(t)}
    \left(\epsilon_l^{(i)}(t) - \epsilon_l^{(i-1)}(t)\right)^2 \dd t
  \neq J^{(i)} - J^{(i-1)}
\end{equation}
for the values $J^{(i)}$, $J^{(i-1)}$ from two consecutive rows of the table.
This is because $\Delta J^{(i)}$ must be evaluated with respect to
a \emph{single} reference field $\epsilon_{l,\text{ref}}^{(i)}(t)$ in
Eq.~\eqref{eq:g_a}, whereas the reported $J^{(i)}$ and $J^{(i-1)}$ use different
reference fields, $\epsilon_{l}^{(i-1)}(t)$ and $\epsilon_{l}^{(i-2)}(t)$
respectively (the guess field in each iteration).

\begin{figure}[tb]
  \centering
  \includegraphics{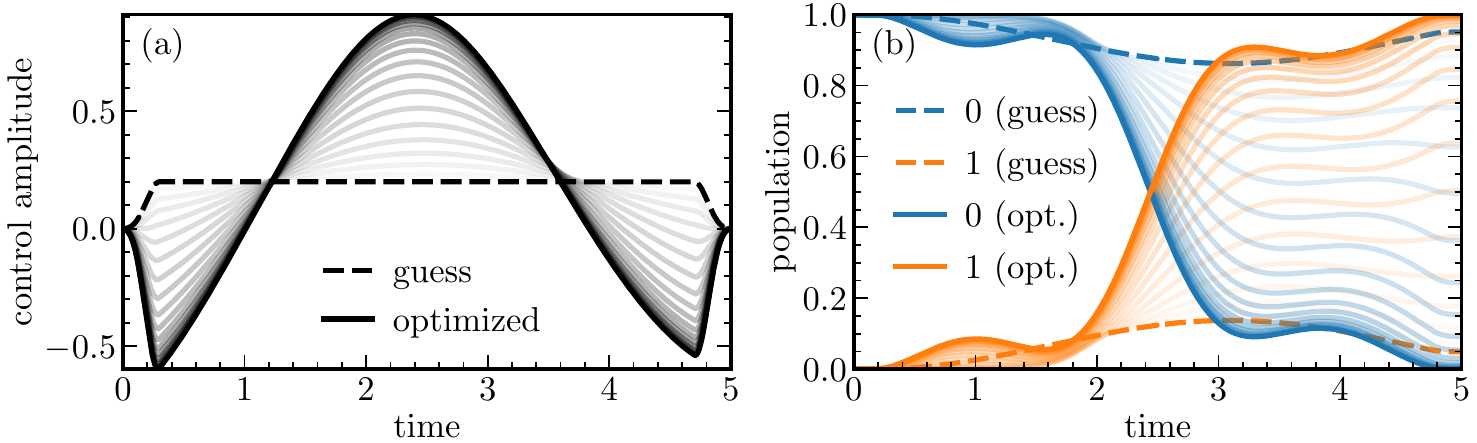}
  \caption{(Color online) Control fields and population dynamics for each
  iteration of the optimization procedure.
  (a) The initial guess control (dashed black line) and optimized controls
  (solid lines), with increasing opacity for each iteration of the optimization.
  The solid black line is the final optimized control.
  (b) The population dynamics for the two levels $\ket{0}$ (blue) and $\ket{1}$
  (orange).  The dynamics under the guess control are indicated by the dashed
  lines, and the dynamics under the optimized control of each iteration by the
  solid lines, with the opacity increasing with the iteration number. All
  quantities are in dimensionless units.
  }%
  \label{fig:tls_example}
\end{figure}
Figure~\ref{fig:tls_example} illustrates the progression of the optimization
procedure.
Panel (a) shows how the control field changes from the initial guess (dashed
line) in each iteration.
All these control fields are available in the \texttt{Result} object due to the
parameter \texttt{store\_all\_pulses} in line~91 of the example.
The optimized control fields are shown with increasing opacity for each
iteration in the optimization.
We can verify that the magnitude of the change in the control field in each
iteration corresponds to the relative magnitude to the values in the column
labeled $\int g_a(t) \dd t$ in the output; as the optimization converges, the
change in each iteration becomes smaller.
All optimized controls preserve the boundary condition of a smooth
switch-on/-off from zero at $t=t_0=0$ and $t=T=5$, due to the choice of the
\texttt{update\_shape}.
Panel (b) shows the population dynamics corresponding to the optimized control
fields, obtained from plugging the optimized controls into the
\texttt{objectives} and propagating the system with a call to the
\texttt{mesolve} method.
Again, the guess is indicated by the dashed line, and the opacity of the solid
lines increases with the iteration number.
We can verify the population transfer of only $0.049$ under the guess control
and the near perfect transfer ($\approx 0.999$) under the final optimized
control.

\section{Common optimization tasks}%
\label{sec:common_optimization_tasks}

In the following, we discuss some of the most common tasks in quantum control
and how they may be realized using the \texttt{krotov} package. The code
snippets in this section are also available as complete examples in the form of
interactive Jupyter notebooks in the Examples section of the online
documentation, see Appendix~\ref{apx:package_docs}.

\subsection{Complex-valued controls}%
\label{sec:task_complex_controls}

When using the rotating wave approximation (RWA), it is important to remember
that the target states are usually defined in the lab frame, not in the rotating
frame.
This is relevant for the construction of \(\ket{\chi_k(T)}\). When doing a
simple optimization, such as a state-to-state or a gate optimization, the
easiest approach is to transform the target states to the rotating frame before
calculating \(\ket{\chi_k(T)}\). This is both straightforward and numerically
efficient.

In the RWA, the control fields are usually complex-valued.
In this case, the Krotov update equation is valid for both the real and the
imaginary part independently.
The most straightforward implementation of the method is to allow for
real-valued controls only, requiring that any control Hamiltonian with a
complex-valued control field is rewritten as two independent control
Hamiltonians, one for the real part and one for the imaginary part of the
control field.
For example,
\begin{equation}
  \epsilon^*(t) \Op{a} + \epsilon(t) \Op{a}^\dagger
  =  \epsilon_{\text{re}}(t) (\Op{a} + \Op{a}^\dagger)
    + \epsilon_{\text{im}}(t) (i \Op{a}^\dagger - i \Op{a})
\end{equation}
with two independent control fields \(\epsilon_{\text{re}}(t)=
\Re[\epsilon(t)]\) and \(\epsilon_{\text{im}}(t) = \Im[\epsilon(t)]\) with the
control Hamiltonian $\Op{a} + \Op{a}^\dagger$ and $i \Op{a}^\dagger - i \Op{a}$,
respectively.

\subsection{Optimization towards a quantum gate}%
\label{sec:task_gate_oct}

To optimize towards a quantum gate \(\Op{O}\) in a \emph{closed} quantum system,
set one \texttt{Objective} for each state in the logical basis, with the basis
state \(\ket{\phi_k}\) as the \texttt{initial\_state} and \(\ket{\phi_k^{\tgt}}
= \Op{O} \ket{\phi_k}\) as the \texttt{target}, cf.~Eq.~\eqref{eq:JTre}.
The helper routine \texttt{gate\_objectives} constructs the appropriate list of
objectives, e.g.\ for a single-qubit Pauli-X gate:
\begin{lstlisting}[language=python, firstline=26, lastline=30]
objectives = krotov.gate_objectives(
    basis_states=[qutip.ket('0'), qutip.ket('1')],
    gate=qutip.operators.sigmax(),
    H=H,
)
\end{lstlisting}

The \texttt{gate\_objectives} routine allows for open quantum systems as well.
The parameter \texttt{liouville\_states\_set} indicates that the system dynamics
are in Liouville space and defines the choice of an appropriate (minimal) set of
matrices to track the optimization~\cite{GoerzNJP2014}.
For example, to optimize for a $\sqrt{\text{iSWAP}}$ gate in an open quantum
system, three appropriately chosen density matrices $\Op{\rho}_1$,
$\Op{\rho}_2$, $\Op{\rho}_3$ are sufficient to track the optimization
progress~\cite{GoerzNJP2014}.
Different emphasis can be put on each matrix, through relative weights 20:1:1 in
the example below:
\begin{lstlisting}[language=python, firstline=28, lastline=34]
objectives = krotov.gate_objectives(
    basis_states=[qutip.ket(l) for l in ['00', '01', '10', '11']],
    gate=qutip.gates.sqrtiswap(),
    H=L,  # Liouvillian super-operator (qutip.Qobj instance)
    liouville_states_set='3states',
    weights=[20, 1, 1],
)
\end{lstlisting}

On many quantum computing platforms, applying arbitrary single-qubit gates is
easy compared to entangling two-qubit gates.
A specific entangling gate like CNOT is combined with single-qubit gates to form
a universal set of gates.
For a given physical system, it can be hard to know a-priori which entangling
gates are easy or even possible to realize.
For example, trapped neutral atoms only allow for the realization of diagonal
two-qubit gates~\cite{JakschPRL2000,GoerzNJP2014} like CPHASE\@.
However, the CPHASE gate is ``locally equivalent'' to CNOT\@: only additional
single-qubit operations are required to obtain one from the other.
A ``local-invariants functional''~\cite{MullerPRA11} defines an optimization
with respect to a such a local equivalence class, and thus is free to find the
specific realization of a two-qubit gate that is easiest to realize.
The objectives for such an optimization are generated by passing
\texttt{local\_invariants=True} to \texttt{gate\_objectives}.

Generalizing the idea further, the relevant property of a gate is often its
entangling power, and the requirement for a two-qubit gate in a universal set of
gates is that it is a ``perfect entangler''.
A perfect entangler can produce a maximally entangled state from a separable
input state.
Since 85\% of all two-qubit gates are perfect
entanglers~\cite{WattsE2013,MuszPRA2013}, a functional that targets an arbitrary
perfect entangler~\cite{WattsPRA2015,GoerzPRA2015} solves the control problem
with the least constraints.
The objectives for this optimization are initialized by passing
\texttt{gate='PE'} to \texttt{gate\_objectives}.
Both the optimization towards a local equivalence class and an arbitrary perfect
entangler may require use of the second-order update equation, see
Sec.~\ref{sec:task_second_order}.

\subsection{Ensemble optimization as a way to ensure robust controls}%
\label{sec:task_ensemble_oct}

Control fields can be made robust with respect to variations in the system by
performing an ``ensemble optimization''~\cite{GoerzPRA2014}.
The idea is to sample a representative selection of possible system
Hamiltonians, and to optimize over an average of the entire ensemble.
In the functional, Eq.~\eqref{eq:functional}, respectively the update
Eq.~\eqref{eq:krotov_first_order_update}, the index $k$ now numbers not only the
states, but also different ensemble Hamiltonians: \(\Op{H}(\{\epsilon_l(t)\})
\rightarrow \{\Op{H}_k(\{\epsilon_l(t)\})\}\).

The example considered in Ref.~\cite{GoerzPRA2014} is that of a CPHASE two-qubit
gate on trapped Rydberg atoms.
Two classical fluctuations contribute significantly to the gate error:
deviations in the pulse amplitude ($\Omega = 1$ ideally), and fluctuations in
the energy of the Rydberg level ($\Delta_{\text{ryd}} = 0$ ideally). We also
take into account decay and dephasing, and thus optimize in Liouville space,
setting the objectives as in Sec.~\ref{sec:task_gate_oct}:
\begin{lstlisting}[language=python, firstline=40, lastline=49]
from math import pi  # standard library
objectives = krotov.gate_objectives(
    basis_states=[
        qutip.ket(l) for l in ['00', '01', '10', '11']
    ],
    gate=qutip.gates.cphase(pi),
    H=L(omega=1, delta=0),
    liouville_states_set='3states',
    weights=[0, 1, 1]
)
\end{lstlisting}
This will result in a list of two objectives
for the density matrices $\Op{\rho}_2$ and $\Op{\rho}_3$ defined in
Ref.~\cite{GoerzNJP2014}.
The state $\Op{\rho}_1$ is omitted by setting its weight to zero, as the target
gate is diagonal.
The function \texttt{L} is assumed to return the Liouvillian for the system with
given values for $\Omega$ and $\Delta_{\text{ryd}}$.

An appropriate set of ensemble objectives (extending the objectives defined
above) can now be generated with the help of the \texttt{ensemble\_objectives}
function.
\begin{lstlisting}[language=python, firstline=51, lastline=59]
import itertools  # standard library
ensemble_liouvillians = [
    L(omega, delta)
    for (omega, delta)
    in itertools.product(omega_vals, delta_vals)
]
objectives = krotov.objectives.ensemble_objectives(
    objectives, ensemble_liouvillians
)
\end{lstlisting}
Here \texttt{omega\_vals} and
\texttt{delta\_vals} is assumed to contain values sampling the space of
perturbed values $\Omega \neq 1$ and $\Delta_{\text{ryd}} \neq 0$.
For $M-1$ \texttt{ensemble\_liouvillians}, i.e.
$M$ systems including the original unperturbed system, the above call results in
a list of $2M$ \texttt{objectives}.
Note that all elements of \texttt{ensemble\_liouvillians} share the same control
pulses.
As shown in Ref.~\cite{GoerzNJP2014}, an optimization over the average of all
these objectives via the functional in Eq.~\eqref{eq:JTre} results in controls
that are robust over a wide range of system perturbations.

\subsection{Optimization of non-linear problems or non-convex functionals}%
\label{sec:task_second_order}

In Refs.~\cite{WattsPRA2015,GoerzPRA2015}, a non-convex final-time functional
for the optimization towards an arbitrary perfect entangler is considered.
In order to guarantee monotonic convergence, the Krotov update equation must be
constructed to second order, see Appendix~\ref{apx:krotov_update_second_order}.
In practice, this means we must specify a scalar function $\sigma(t)$ that
serves as the coefficient to the second order contribution.

For this specific example, a suitable choice is
\begin{equation}
  \sigma(t) \equiv -\max\left(\varepsilon_A,2A+\varepsilon_A\right)\,,
\end{equation}
where \(\varepsilon_A\) is a small non-negative number.
The optimal value for \(A\) in each iteration can be approximated numerically
as~\cite{ReichJCP12}
\begin{equation}
  \label{eq:numericalA}
  A = \frac{
    \sum_{k=1}^{N} 2 \Re\left[
      \langle \chi_k(T) \vert \Delta\phi_k(T) \rangle \right] + \Delta J_T}{
    \sum_{k=1}^{N} \Abs{\Delta\phi_k(T)}^2} \,,
\end{equation}
with
\begin{equation}
  \Delta J_T \equiv J_T(\{\phi_k^{(i)}(T)\}) -J_T(\{\phi_k^{(i-1)}(T)\})\,.
\end{equation}

In the \texttt{krotov} package, in order to make use of the second order
contribution to the pulse update, we pass a parameter \texttt{sigma} to the
\texttt{optimize\_pulses} function:
\begin{lstlisting}[language=python, numbers=left, firstline=108, lastline=142]
class sigma(krotov.second_order.Sigma):
    def __init__(self, A, epsA=0):
        self.A = A
        self.epsA = epsA

    def __call__(self, t):
        return -max(self.epsA, 2 * self.A + self.epsA)

    def refresh(
        self, forward_states, forward_states0,
        chi_states, chi_norms, optimized_pulses,
        guess_pulses, objectives, result,
    ):
        try:
            # info_vals contains values of PE functional
            Delta_J_T = (
                result.info_vals[-1][0] - result.info_vals[-2][0]
            )
        except IndexError:  # first iteration
            Delta_J_T = 0
        self.A = krotov.second_order.numerical_estimate_A(
            forward_states, forward_states0, chi_states,
            chi_norms, Delta_J_T
        )


oct_result = krotov.optimize_pulses(
    objectives,
    pulse_options=pulse_options,
    tlist=tlist,
    propagator=krotov.propagators.expm,
    chi_constructor=chi_constructor,  # from weylchamber package
    info_hook=calculate_PE_val,
    sigma=sigma(A=0.0),
)
\end{lstlisting}
The function \texttt{krotov.second\_order.numerical\_estimate\_A} implements
Eq.~\eqref{eq:numericalA}.
The function defined by the instantiated \texttt{sigma} is used for the pulse
update, and then the internal parameter, $A$ in this case, is automatically
updated at the end of each iteration, via the \texttt{sigma}'s \texttt{refresh}
method.

Even when the second order update equation is mathematically required to
\emph{guarantee} monotonic convergence, often an optimization with the
first-order update equation~\eqref{eq:krotov_first_order_update} will give
converging results.
Since the second order update requires more numerical resources (calculation and
storage of the states \(\ket{\Delta\phi_k(t)}\), see
Appendix~\ref{apx:pseudocode}), it is advisable to attempt an optimization with
the first-order update equation first, and to only use the second order when the
first order proves insufficient.

\section{Comparison of Krotov's method and other optimization methods}%
\label{sec:comparison}

In the following, we compare Krotov's method to other numerical optimization
methods that have been used widely in quantum control, with an emphasis on
methods that have been implemented as open source software.
We first discuss iterative schemes derived from general variational calculus in
Section~\ref{sec:comparison_variational_calculus} before making the connection
to Krotov's method in particular, in Section~\ref{sec:comparison_krotov}.
We then compare with GRadient Ascent Pulse Engineering
(GRAPE)~\cite{GlaserEPJD2015,KhanejaJMR05} in
Section~\ref{sec:comparison_grape}, before highlighting the differences with
gradient-free methods in Section~\ref{sec:comparison_gradientfree}.
Finally, Section~\ref{sec:comparison_choose_method} provides some guidance for
the choice of an appropriate optimization method for particular circumstances.

\subsection{Iterative schemes from variational calculus}
\label{sec:comparison_variational_calculus}

Gradient-based optimal control methods derive the condition for the optimal
control field from the application of the variational principle to the
optimization functional in Eq.~\eqref{eq:functional}.
Since the functional depends both on the states and the control field, it is
necessary to include the equation of motion (Schrödinger or
Liouville-von-Neumann) as a constraint.
That is, the states $\{\ket{\phi_k}\}$ must be compatible with the equation of
motion under the control fields $\{\epsilon_l(t)\}$.
In order to convert the constrained optimization problem into an unconstrained
one, the equation of motion is included in the functional with the co-states
$\ket{\chi_k(t)}$ as  Lagrange
multipliers~\cite{KosloffCP1989,ShiJCP1990,ShiCPC1991,Tannor91}.

The necessary condition for an extremum becomes $\delta J = 0$ for this extended
functional.
Evaluation of the extremum condition results in~\cite{Tannor91}
\begin{equation}
  \label{eq:variational_update}
  \Delta \epsilon_l(t) \propto \frac{\delta J}{\delta \epsilon_l} \propto
    \Im \big\langle
    \chi_k(t)
    \big\vert
    \Op{\mu}
    \big\vert
    \phi_k(t)
    \big\rangle\,,
\end{equation}
where $\Op{\mu} = \partial \Op{H} / \partial \epsilon_l(t)$ is the operator coupling
to the field $\epsilon_l(t)$.
Equation~\eqref{eq:variational_update} is both continuous in time and implicit
in $\epsilon_l(t)$ since the states $\ket{\phi_k(t)}$, $\ket{\chi_k(t)}$ also
depend on $\epsilon_l(t)$.
Numerical solution of Eq.~\eqref{eq:variational_update} thus requires an
iterative scheme and a choice of time discretization.

The most intuitive time-discretization yields a \emph{concurrent} update
scheme~\cite{Tannor91,Tannor92,SomloiCP1993},
\begin{equation}
  \label{eq:concurrent_update}
  \Delta \epsilon_l^{(i)}(t) \propto
    \Im \big\langle
    \chi_k^{(i-1)}(t)
    \big\vert
    \Op{\mu}
    \big\vert
    \phi_k^{(i-1)}(t)
    \big\rangle\,.
\end{equation}
Here, at iterative step $(i)$, the backward-propagated co-states
$\{\ket{\chi_k(t)}\}$ and the forward-propagated states  $\{\ket{\phi_k(t)}\}$
both evolve under the 'guess' controls $\epsilon^{(i-1)}_l(t)$ of that
iteration.
Thus, the update is determined entirely by information from the previous
iteration and can be evaluated at each point $t$ independently.
However, this scheme does not guarantee monotonic convergence, and requires a
line search to determine the appropriate magnitude of the pulse
update~\cite{Tannor91}.

A further ad-hoc modification of the functional~\cite{ZhuJCP98} allows to
formulate a family of update schemes that do guarantee monotonic
convergence~\cite{MadayJCP2003,OhtsukiJCP2004}.
These schemes introduce separate fields $\{\epsilon_l(t)\}$ and
$\{\tilde\epsilon_l(t)\}$ for the  forward and backward propagation,
respectively, and use the update scheme~\cite{WerschnikJPB2007}
\begin{subequations}
  \label{eq:zhu_rabitz_update}
  \begin{align}
    \epsilon_l^{(i)}(t)
      & = (1-\delta) \tilde\epsilon_l^{(i-1)}(t) - \frac{\delta}{\alpha}
      \Im \big\langle
        \chi_k^{(i-1)}(t)
        \big\vert
        \Op{\mu}
        \big\vert
        \phi_k^{(i)}(t)
        \big\rangle \\
    \tilde\epsilon_l^{(i)}(t)
      & = (1-\eta) \epsilon_l^{(i-1)}(t) - \frac{\eta}{\alpha}
      \Im \big\langle
        \chi_k^{(i)}(t)
        \big\vert
        \Op{\mu}
        \big\vert
        \phi_k^{(i)}(t)
        \big\rangle\,,
  \end{align}
\end{subequations}
with $\delta, \eta \in [0, 2]$  and an arbitrary step width $\alpha$.
For the control of wavepacket dynamics, an implementation of this generalized
class of algorithms is available in the WavePacket Matlab
package~\cite{SchmidtCPC2018}.

\subsection{Krotov's method}%
\label{sec:comparison_krotov}

The method developed by
Krotov~\cite{KrotovEC1983,KrotovCC1988,Krotov.book,KonnovARC99} and later
translated to the language of quantum control by Tannor and
coworkers~\cite{Tannor92,SomloiCP1993,BartanaJCP1997,SklarzPRA2002,ReichJCP12}
takes a somewhat unintuitive approach to disentangle the interdependence of
field and states by adding a zero to the functional.
This allows to \emph{construct} an updated control field that is guaranteed to
lower the value of the functional, resulting in monotonic convergence.
The full method is described in Appendix~\ref{apx:krotov_update}, but its
essence can be boiled down to the update in each iteration $(i)$,
Eq.~\eqref{eq:update}, taking the form
\begin{equation}
  \label{eq:sequential_update}
  \Delta \epsilon_l^{(i)}(t) \propto
    \Im \big\langle
    \chi_k^{(i-1)}(t)
    \big\vert
    \Op{\mu}
    \big\vert
    \phi_k^{(i)}(t)
    \big\rangle\,,
\end{equation}
with co-states $\ket{\chi_k(t)^{(i-1)}}$ backward-propagated under the
\emph{guess} controls $\{\epsilon_l^{(i-1)}(t)\}$ and the states
$\ket{\phi^{(i)}_k(t)}$ forward-propagated under the \emph{optimized} controls
$\{\epsilon_l^{(i)}(t)\}$.
Compared to the \emph{concurrent} form of Eq.~\eqref{eq:concurrent_update}, the
Krotov update scheme is \emph{sequential}: The update at time $t$ depends on the
states forward-propagated using the updated controls at all previous times, see
Appendix~\ref{apx:krotov_update_discretization} for details.

It is worth noting that the sequential update can be recovered as a limiting
case of the monotonically convergent class of algorithms in
Eq.~\eqref{eq:zhu_rabitz_update}, for $\delta = 1$, $\eta = 0$.
This may explain why parts of the quantum control community consider \emph{any}
sequential update scheme as ``Krotov's
method''~\cite{SchirmerNJP2011,MachnesPRA11}.
However, following Krotov's
construction~\cite{KrotovEC1983,KrotovCC1988,Krotov.book,KonnovARC99} requires
no ad-hoc modification of the functional and can thus be applied more generally.
In particular, as discussed in Section~\ref{sec:task_second_order} and
Appendix~\ref{apx:krotov_update_second_order}, a second-order construction can
address non-convex functionals.

In all its
variants~\cite{Tannor92,SomloiCP1993,BartanaJCP1997,SklarzPRA2002,ReichJCP12},
Krotov's method is a first-order gradient with respect to the control fields
(even in the second-order construction which is second order only with respect
to the states). As the optimization approaches the optimum, this gradient can
become very small, resulting in slow convergence.
It is possible to extend Krotov's method to take into account information from
the quasi-Hessian~\cite{EitanPRA11}.
However, this ``K-BFGS'' variant of Krotov's method is a substantial extension
to the procedure as described in Appendix~\ref{apx:pseudocode}, and is currently
not supported by the \texttt{krotov} package.

The update Eq.~\eqref{eq:sequential_update} is specific to the running cost in
Eq.~\eqref{eq:g_a}.
In most of the schemes derived from variational calculus, cf.
Section~\ref{sec:comparison_variational_calculus}, a constraint on the
\emph{pulse fluence} is used instead.
Formally, this is also compatible with Krotov's method, by choosing
$\epsilon_{l, \text{ref}}^{(i)}(t) \equiv 0$ in
Eq.~\eqref{eq:g_a}~\cite{PalaoPRL2002}.
It turns the \emph{update}
equations~(\ref{eq:sequential_update},~\ref{eq:concurrent_update}) into
\emph{replacement} equations, with $\epsilon_l^{(i)}(t)$ on the left-hand side
instead of $\Delta\epsilon_l^{(i)}(t)$, cf.~Eq.~\eqref{eq:zhu_rabitz_update} for
$\delta = 1$, $\eta = 0$.
In our experience, this leads to numerical instability and should be avoided.
A mixture of \emph{update} and \emph{replacement} is possible when a penalty of
the pulse fluence is necessary~\cite{GoerzPhd2015}.

\subsection{GRadient Ascent Pulse Engineering (GRAPE)}%
\label{sec:comparison_grape}

While the monotonically convergent methods based on variational calculus must
``guess'' the appropriate time discretization, and Krotov's method finds the
sequential time discretization by a clever construction, the GRAPE method
sidesteps the problem by discretizing the functional \emph{first}, before
applying the variational calculus.

Specifically, we consider the piecewise-constant discretization of the dynamics
onto a time grid, where the final time states $\{\ket{\phi_k^{(i-1)}(T)}\}$
resulting from the time evolution of the initial states $\{\ket{\phi_k}\}$ under
the guess controls $\epsilon^{(i-1)}_n$ in iteration $(i)$ of the optimization
are obtained as
\begin{equation}%
\label{eq:discrete_time_evolution} \ket{\phi^{(i-1)}_k(T)} = \Op{U}^{(i-1)}_{N_T} \dots
\Op{U}^{(i-1)}_{n} \dots \Op{U}^{(i-1)}_{1} \big\vert \phi_k \big\rangle\,,
\end{equation}
where  $\Op{U}^{(i-1)}_{n}$ is the time evolution operator on the time interval
$n$ in Hilbert space,
\begin{equation}
\Op{U}^{(i-1)}_{n} = \exp\Bigg[ -\frac{\mathrm{i}}{\hbar} \Op{H}\big(
\underbrace{\epsilon^{(i-1)}(\tilde t_{n-1})}_{\epsilon^{(i-1)}_n} \big) \dd
t\Bigg];\qquad \tilde{t}_n \equiv t_n + \dd t / 2\,.
\end{equation}
The independent control parameters are now the scalar values $\epsilon_n$,
respectively $\epsilon_{ln}$ if there are multiple control fields indexed by
$l$.

The GRAPE method looks at the direct gradient $\partial J/\partial \epsilon_n$
and updates each control parameter in the direction of that
gradient~\cite{KhanejaJMR05}.
The step width must be determined by a line search.

Typically, only the final time functional $J_T$ has a nontrivial gradient.
For simplicity, we assume that $J_T$ can be expressed in terms of
the complex overlaps $\{\tau_k\}$ between the target states
\(\{\ket{\phi_k^{\tgt}}\}\) and the propagated states
\(\{\ket{\phi_k(T)}\}\), as e.g.\ in Eqs.~(\ref{eq:JTss},~\ref{eq:JTre}).
Using Eq.~\eqref{eq:discrete_time_evolution} leads to
\begin{equation}%
  \label{eq:gradient}
  \begin{split}
    \frac{\partial \tau_k}{\partial \epsilon_n} &= \frac{\partial}{\partial
    \epsilon_n} \big\langle \phi_k^{\tgt} \big\vert \Op{U}^{(i-1)}_{N_T} \dots
    \Op{U}^{(i-1)}_{n} \dots \Op{U}^{(i-1)}_{1} \big\vert \phi_k \big\rangle \\
    &=
        \underbrace{%
    \big\langle \phi_k^{\tgt} \big\vert
              \Op{U}^{(i-1)}_{N_T} \dots \Op{U}^{(i-1)}_{n+1}}_{%
    \bra{\chi^{(i-1)}_k(t_{n+1})} } \,
    \frac{\partial\Op{U}^{(i-1)}_{n}}{\partial\epsilon_n} \,
         \underbrace{%
    \Op{U}^{(i-1)}_{n-1} \dots \Op{U}^{(i-1)}_{1} \big\vert
            \phi_k \big\rangle}_{%
    \ket{\phi^{(i-1)}_k(t_n)} }\
  \end{split}
\end{equation}
as the gradient of these overlaps.
The gradient for $J_T$, respectively $J$ if there
are additional running costs then follows from the chain rule.
The numerical evaluation of Eq.~\eqref{eq:gradient} involves the
backward-propagated states \(\ket{\chi_k(t_{n+1})}\) and the
forward-propagated states \(\ket{\phi_k(t_n)}\).
As only states from iteration $(i-1)$ enter in the gradient, GRAPE is
a \emph{concurrent} scheme.

The comparison of the sequential update equation~\eqref{eq:sequential_update} of
Krotov's method and the concurrent update equation~\eqref{eq:concurrent_update}
has inspired a sequential evaluation of the ``gradient'', modifying the
right-hand side of Eq.~\eqref{eq:gradient} to $\langle \chi_k^{(i-1)}(t_{n+1})
\vert \partial_\epsilon U_n^{(i-1)} \vert \phi_k^{(i)}(t_n)\rangle$.
That is, the states $\{\ket{\phi_k(t)}\}$ are forward-propagated under the
optimized field~\cite{SchirmerJMO2009}.
This can be generalized to ``hybrid'' schemes that interleave concurrent and
sequential calculation of the gradient~\cite{MachnesPRA11}.
An implementation of the concurrent/sequential/hybrid gradient is available in
the DYNAMO Matlab package~\cite{MachnesPRA11}.
The sequential gradient scheme is sometimes referred to as
``Krotov-type''~\cite{MachnesPRA11, FloetherNJP12}.
To avoid confusion with the specific method defined in
Appendix~\ref{apx:krotov_update}, we prefer the name ``sequential GRAPE''.

GRAPE does not give a guarantee of monotonic convergence.
As the optimization approaches the minimum of the functional, the first order
gradient is generally insufficient to drive the optimization
further~\cite{EitanPRA11}.
To remedy this, a numerical estimate of the Hessian \(\partial^2 J_T/\partial
\epsilon_j \partial \epsilon_{j^\prime}\) should also be included in the
calculation of the update.
The L-BFGS-B quasi-Newton method~\cite{ByrdSJSC1995,ZhuATMS97} is most commonly
used for this purpose, resulting in the ``Second-order
GRAPE''~\cite{FouquieresJMR2011} or ``GRAPE-LBFGS'' method.
L-BFGS-B is implemented as a Fortran library~\cite{ZhuATMS97} and widely
available, e.g.\ wrapped in optimization toolboxes like SciPy~\cite{Scipy}.
This means that it can be easily added as a ``black box'' to an existing
gradient optimization.
As a result, augmenting GRAPE with a quasi-Hessian is essentially ``for free''.
Thus, we always mean GRAPE to refer to GRAPE-LBFGS\@.
Empirically, GRAPE-LBFGS \emph{usually} converges monotonically.

Thus, for (discretized) time-continuous controls, both GRAPE and
Krotov's method can generally be used interchangeably. Historically, Krotov's
method has been used primarily in the control of molecular dynamics, while
GRAPE has been popular in the NMR community.
Some potential benefits of Krotov's method compared to GRAPE are~\cite{EitanPRA11}:
\begin{itemize}
  \item Krotov's method mathematically guarantees monotonic convergence in the
    continuous-time limit. There is no line search required for the step width
    $1 / \lambda_{a, l}$.
  \item The sequential nature of Krotov's update scheme, with information from
    earlier times entering the update at later times within the same iteration,
    results in faster convergence than the concurrent update in
    GRAPE~\cite{MachnesPRA11,JaegerPRA14}.  This advantage disappears as the
    optimization approaches the optimum~\cite{EitanPRA11}.
  \item The choice of functional $J_T$ in Krotov's method only enters in the
    boundary condition for the backward-propagated states,
    Eq.~\eqref{eq:chi_boundary}, while the update equation stays the same
    otherwise.
    In contrast, for functionals $J_T$ that do not depend trivially on the
    overlaps~\cite{NevesJMR2009,NguyenJMR2017,AnselPRA2017,SpindlerJMR2012,TosnerACIE2018},
    the evaluation of the gradient in GRAPE may deviate significantly from its
    usual form, requiring a problem-specific implementation from scratch.  This
    may be mitigated by the use of automatic differentiation in future
    implementations~\cite{LeungPRA2017,AbdelhafezPRA2019}.
\end{itemize}

GRAPE has a significant advantage if the controls are not time-continuous, but
are \emph{physically} piecewise constant (``bang-bang control''). The
calculation of the GRAPE-gradient is unaffected by this, whereas Krotov's method
can break down when the controls are not approximately continuous.
QuTiP contains an implementation of GRAPE limited to this use case.

Variants of gradient-ascent can be used to address \emph{pulse
parametrizations}.
That is, the control parameters may be arbitrary parameters of the control field
(e.g., spectral coefficients) instead of the field amplitude $\epsilon_n$ in a
particular time interval.
This is often relevant to design control fields that meet experimental
constraints.
One possible realization  is to calculate the gradients for the control
parameters from the gradients of the time-discrete control amplitudes via the
chain rule~\cite{WinckelIP2008,SkinnerJMR2010,MotzoiPRA2011,LucarelliPRA2018}.
This approach has recently been named  ``GRadient Optimization Using
Parametrization'' (GROUP)~\cite{SorensenPRA2018}.
An implementation of several variants of GROUP is available in the QEngine C++
library~\cite{SorensenCPC2019}.
An alternative for a moderate number of control parameters is
``gradient-optimization of analytic controls'' (GOAT)~\cite{MachnesPRL2018}.
GOAT evaluates the relevant gradient with forward-mode differentiation; that is,
${\partial \tau_k}/{\partial \epsilon_n}$ is directly evaluated alongside
$\tau_k$.
For $N = \Abs{\{\epsilon_m\}}$ control parameters, this implies $N$ forward
propagations of the state-gradient pair per iteration.
Alternatively, the $N$ propagations can be concatenated into a single
propagation in a Hilbert space enlarged by a factor $N$ (the original state
paired with $N$ gradients).

A benefit of GOAT over the more general GROUP is that it does not piggy-back on
the piecewise-constant discretization of the control field, and thus may avoid
the associated numerical error.
This allows to optimize to extremely high fidelities as required for some error
correction protocols~\cite{MachnesPRL2018}.
\subsection{Gradient-free optimization}%
\label{sec:comparison_gradientfree}

In situations where the problem can be reduced to a relatively small number of
control parameters (typically less than $\approx20$, although this number may be
pushed to $\approx50$ by sequential increase of the number of parameters and
re-parametrization~\cite{RachPRA2015,GoetzPRA2016}), gradient-free optimization
becomes feasible.
The most straightforward use case are controls with an analytic shape (e.g.
due to the constraints of an experimental setup), with just a few free
parameters.
As an example, consider control pulses that are restricted to a Gaussian shape,
so that the only free parameters are peak amplitude, pulse width and delay.
The control parameters are not required to be parameters of a time-dependent
control, but may also be static parameters in the Hamiltonian, e.g.\ the
polarization of the laser beams utilized in an experiment~\cite{HornNJP2018}.

A special case of gradient-free optimization is the Chopped RAndom Basis (CRAB)
method~\cite{DoriaPRL11,CanevaPRA2011}.
The essence of CRAB is in the specific choice of the parametrization in terms of
a low-dimensional \emph{random} basis, as the name implies.
Thus, it can be used when the parametrization is not pre-defined as in the case
of direct free parameters in the pulse shape discussed above.
The optimization itself is normally performed by Nelder-Mead simplex based on
this parametrization, although any other gradient-free method could be used as
well.
An implementation of CRAB is available in QuTiP. CRAB is prone to getting stuck
in local minima of the optimization landscape.
To remedy this, a variant of CRAB, ``dressed CRAB'' (DCRAB) has been
developed~\cite{RachPRA2015} that re-parametrizes the controls when this
happens.

Gradient-free optimization does not require backward propagation, only forward
propagation of the initial states and evaluation of the optimization functional
$J$.
The functional is not required to be analytic.
It may be of a form that does not allow calculation of the gradients \(\partial
J_T / \partial \bra{\phi_k}\) (Krotov's method) or \(\partial J / \partial
\epsilon_j\) (GRAPE). The optimization also does not require  any storage of
states.
However, the number of iterations can grow extremely large, especially with an
increasing number of control parameters.
Thus, an optimization with a gradient-free method is not necessarily more
efficient overall compared to a gradient-based optimization with much faster
convergence.
For only a few parameters, however, it can be highly efficient.
This makes gradient-free optimization useful for ``pre-optimization'', that is,
for finding guess controls that are then further optimized with a gradient-based
method~\cite{GoerzEPJQT2015}.

Generally, gradient-free optimization can be easily realized directly in QuTiP
or any other software package for the simulation of quantum dynamics:
\begin{itemize}
  \item
    Write a function that takes an array of optimization parameters as input and
    returns a figure of merit.  This function would, e.g., construct a numerical
    control pulse from the control parameters, simulate the dynamics using {\tt
    qutip.mesolve.mesolve}, and evaluate a figure of merit (like the overlap
    with a target state).
  \item
    Pass the function to {\tt scipy.optimize.minimize} for gradient-free
    optimization.
\end{itemize}
The implementation in {\tt scipy.optimize.minimize} allows to choose between
different optimization methods, with Nelder-Mead simplex being the default.
There exist also more advanced optimization methods available in packages like
NLopt~\cite{NLOpt} or Nevergrad~\cite{nevergrad} that may be worth exploring for
improvements in numerical efficiency and additional functionality such as
support for non-linear constraints.

\subsection{Choosing an optimization method}%
\label{sec:comparison_choose_method}

\begin{figure}[tb]
\centering \includegraphics[width=\textwidth]{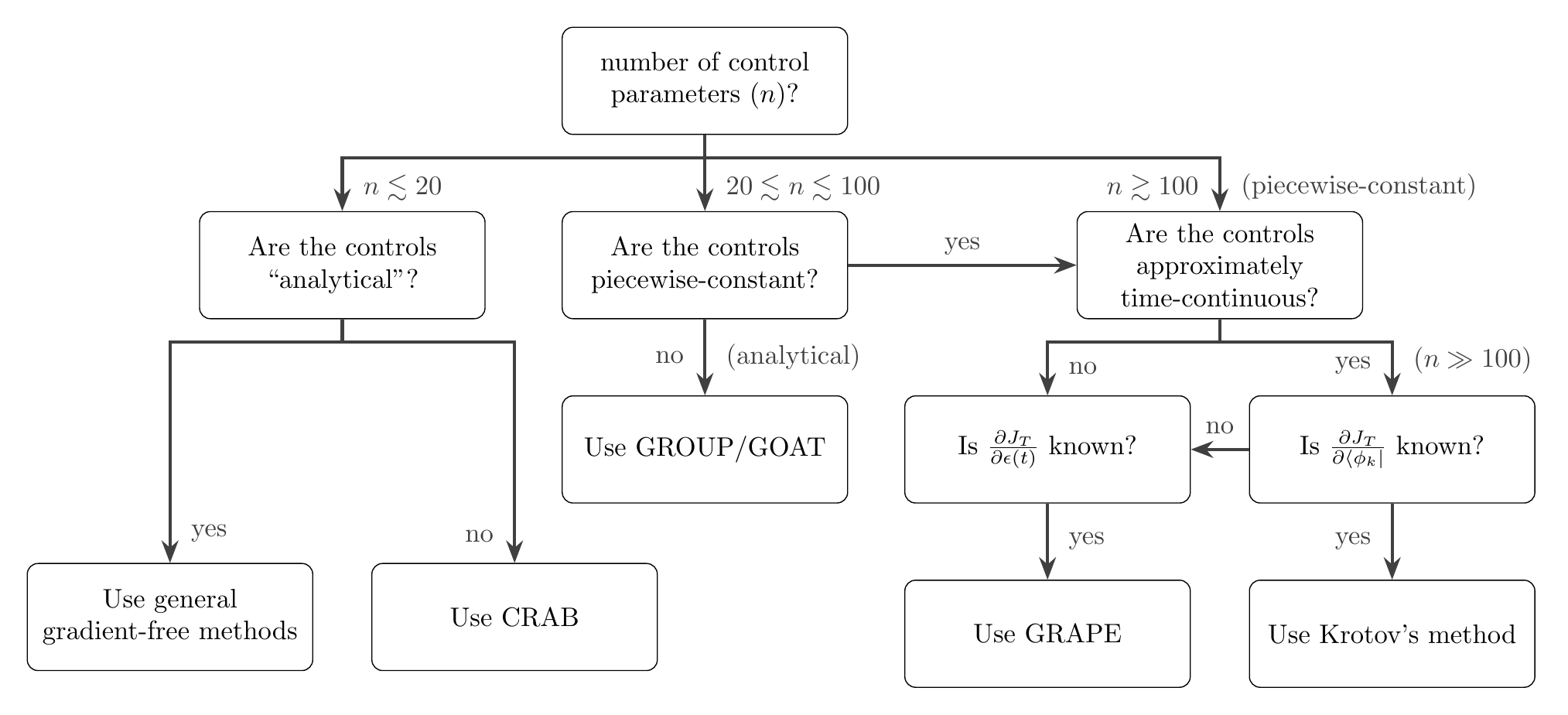}
  \caption{Decision tree for the choice of a numerical open-loop optimization
  method.
  The choice of control method is most directly associated with the number
  of control parameters ($n$).
  For ``piecewise-constant controls'', the control parameters are the values of
  the control field in each time interval. For ``analytical'' controls, we
  assume that the control fields are described by a fixed
  analytical formula parametrized by the control parameters.
  The ``non-analytical'' controls for CRAB refer to
  the \emph{random} choice of a fixed number of spectral components, where the
  control parameters are the coefficients for those spectral components.
  Each method in the diagram is meant to include all its variants, a multitude
  of gradient-free methods and e.g. DCRAB for CRAB, GRAPE-LBFGS and
  sequential/hybrid gradient-descent for GRAPE, and K-BFGS for Krotov's method,
  see text for detail.
  }%
\label{fig:octdecisiontree}
\end{figure}

In the following, we discuss some of the concerns in the choice of optimization
methods.
The discussion is limited to iterative open-loop methods, where the optimization
is based on a numerical simulation of the dynamics.
It excludes analytical control methods such as geometric control, closed-loop
methods, or coherent feedback control; see Ref.~\cite{PetersenIETCTA2010} for an
overview.

Whether to use a gradient-free optimization method, GRAPE, or Krotov's method
depends on the size of the problem, the requirements on the control fields, and
the mathematical properties of the optimization functional.
Gradient-free methods should be used if the number of independent control
parameters is smaller than $\approx 20$, or the functional is of a form that
does not allow to calculate gradients easily.
It is always a good idea to use a gradient-free method to obtain improved guess
pulses for use with a gradient-based method~\cite{GoerzEPJQT2015}.

GRAPE or its variants should be used if the control parameters are discrete,
such as on a coarse-grained time grid, and the derivative of $J$ with respect to
each control parameter is easily computable.
Note that the implementation provided in QuTiP is limited to state-to-state
transitions and quantum gates, even though the method is generally applicable to
a wider range of objectives.

When the control parameters are general analytic coefficients instead of
time-discrete amplitudes, the GROUP~\cite{SkinnerJMR2010,
MotzoiPRA2011,SorensenPRA2018} or GOAT~\cite{MachnesPRL2018} variant of
gradient-ascent may be a suitable choice.
GOAT in particular can avoid the numerical error associated with time
discretization.
However, as the method scales linearly in memory and/or CPU with the number of
control parameters, this is best used when then number of parameters is below
100.

Krotov's method should be used if the control is close to time-continuous, and
if the derivative of \(J_T\) with respect to the states,
Eq.~\eqref{eq:chi_boundary}, can be calculated.
When these conditions are met, Krotov's method gives excellent convergence.
The general family of monotonically convergent iteration
schemes~\cite{MadayJCP2003} may also be used.

The decision tree in Fig.~\ref{fig:octdecisiontree} can guide the choice of an
optimization method.
The key deciding factors are the number of control parameters ($n$) and whether
the controls are time-discrete.
Of course, the parametrization of the controls is itself a choice.
Sometimes, experimental constraints only allow controls that depend on a small
number of tunable parameters.
However, this necessarily limits the exploration of the full physical
optimization landscape.
At the other end of the spectrum, arbitrary time-continuous controls such as
those assumed in Krotov's method have no inherent constraints and are especially
useful for more fundamental tasks, such as mapping the design landscape of a
particular system~\cite{GoerzNPJQI2017} or determining the quantum speed limit,
i.e., the minimum time in which the system can reach a given
target~\cite{CanevaPRL09,GoerzJPB11,SorensenNat16}.

\section{Future perspectives}%
\label{sec:future_perspectives}

While the present implementation of the \texttt{krotov} Python package already
provides the user with the capability to tackle a broad range of optimization
targets in quantum optimal control, possible future additions could enhance its
versatility even further.
A first most welcome extension concerns the capability to parametrize the pulse.
This would allow to guarantee positivity of the control field when optimizing,
e.g., Rabi frequencies instead of pulse amplitudes, or provide a straightforward
way to impose an upper bound $\epsilon_0$ on the field amplitude.
The latter could be achieved, for example, by way of defining
$\epsilon(t)=\epsilon_0 \tanh^2\left(u(t)\right)$~\cite{MuellerPRA2011}.
The simplest approach to adapt the algorithm to such parametrizations is to
consider the Hamiltonian / Liouvillian as a function of $u(t)$ instead of
$\epsilon(t)$.
Then, the update equation will also be formulated with respect to $u(t)$ and
once the optimization is completed the physical pulse $\epsilon(t)$ can be
obtained by direct evaluation.
A caveat in this approach is the fact that the Hamiltonian / Liouvillian will
not be a linear function of $u(t)$ even if it was linear with respect to
$\epsilon(t)$.
As such, additional care needs to be taken regarding the choice of a
sufficiently large value for the inverse step size $\lambda_a$ to preserve
monotonic convergence~\cite{ReichJCP12}.

A second feature worthwhile to add in a future version of the \texttt{krotov}
Python package are state-dependent constraints
$g_b\neq0$~\cite{ReichJCP12,PalaoPRA2008}.
This would enable to optimization towards time-dependent
targets~\cite{KaiserJCP2004, SerbanPRA2005}.
If the constraint is a non-convex function of the states, usage of the
second-order contribution, $\sigma(t)\neq 0$, in the Krotov update
equation~\eqref{eq:krotov_second_order_update} is required to ensure monotonic
convergence.
In this case, $\sigma(t)\neq 0$ is linearly time-dependent~\cite{ReichJCP12}.
The presence of a state-dependent constraint also implies a source term in the
equation of motion for the adjoint states, cf.~Eq.~\eqref{eq:bw_eqm}.
Although this source term may pose some numerical challenges for differential
equation solvers, it should be noted that the solution of a linear Schrödinger
equation with a source term already allows for solving Schrödinger equations
with a general nonlinearity~\cite{SchaeferJCompP2017}.
Assuming an appropriate solver was available, the \texttt{krotov} package would
have to calculate the appropriate source term and pass it to that solver.

Finally, the current implementation of the package does not yet allow for
imposing spectral constraints in the optimization functional, although this is
in principle possible in Krotov's method~\cite{PalaoPRA13,ReichJMO14}.
At first glance, it may be surprising that a method that updates the control
sequentially (time-locally) can include spectral (time-global) constraints
without breaking monotonic convergence.
The key insight is to generalize $g_a(\epsilon(t))$, Eq.~\eqref{eq:g_a}, to a
time-non-local form,
\begin{equation}
  g_a(\epsilon(t),t) = \int_0^T \Delta\epsilon(t) K(t-t^\prime)
  \Delta\epsilon(t^\prime) \dd t^\prime\,.
\end{equation}
Provided the kernel $K(\tau)$ encoding the spectral constraint via a Fourier
transform is positive semi-definite, Krotov's method yields a monotonically
converging optimization algorithm~\cite{ReichJMO14}.
However, the price to pay is the need to solve a Fredholm equation of the second
kind, which has not yet been implemented numerically.
It should be noted that the current version of the \texttt{krotov} package
already supports a less rigorous method to limit the spectral width of the
optimized controls, by applying a simple spectral filter after each iteration.
By mixing the unfiltered and filtered controls, monotonic convergence can be
preserved~\cite{LapertPRA09}.

The above mentioned features concern direct extensions of Krotov's method that
have already been reported in the literature.
Beyond that, Krotov's method could also be combined with other optimization
approaches to overcome some of its inherent limitations.
The most severe limitations are that Krotov's method requires analytically
computable derivatives, see Eq.~\eqref{eq:chi_boundary}, and it searches only in
the local region of the point in the optimization landscape where that
derivative is being evaluated (as any gradient-based method does). The optimized
pulse thus depends on the guess pulse from which the optimization starts.
If the pulse can be parametrized with only a few relevant parameters, the search
can be effectively globalized by scanning those parameters~\cite{GoerzNPJQI17}.
This approach becomes more efficient when pre-optimizing the parameters with a
gradient-free method~\cite{GoerzEPJQT2015}.
In this respect, it will be worthwhile to combine the \texttt{krotov} package
with the nonlinear optimization toolbox NLopt~\cite{NLOpt} containing several
global search methods.
This should not only improve the convergence of the pre-optimization compared to
using the simplex method~\cite{GoerzEPJQT2015} but would, moreover, also allow
for simultaneously optimizing time-dependent and time-independent controls.
The inherent limitation of requiring computable derivatives might be lifted by
combining Krotov's method with automatic differentiation, similar to what has
been achieved for gradient-based optimization in the spirit of
GRAPE~\cite{LeungPRA2017,AbdelhafezPRA2019}.
Finally, it would also be interesting to analyze optimizations using Krotov's
method with machine learning techniques~\cite{BukinPRX2018}.

\section{Conclusions}
\label{sec:conclusions}

We have presented the Python implementation of Krotov's method for quantum
optimal control that comes with a number of example use cases, suitable in
particular for applications in quantum information science.
The hallmark of Krotov's method is fast initial convergence, monotonicity and
aptitude for time-continuous controls.

The \texttt{krotov} package adds to the available tools for optimal control
around the popular Quantum Toolbox in Python (QuTiP). The QuTiP package itself
contains routines for gradient-free optimization and gradient-ascent, currently
limited to state-to-state transitions or quantum gates and to a coarse time
grid.
Our package provides an interface for formulating quantum control problems that
lifts these limitations and aims to be sufficiently general to describe
\emph{any} problem in quantum control.
In future work, the same interface may be used to drive optimization methods
beyond Krotov's method, enabling direct comparison of different methods.

We have given an overview of the most important gradient-free and gradient-based
methods that have been developed thus far.
Each method has its own strengths and weaknesses under different constraints.
Krotov's method in particular excels at finding the \emph{least} constrained
control fields and is thus particularly useful for exploring the fundamental
limits of control in a given quantum system.
On the other hand, when there are in fact strong external constraints on the
controls due to experimental limitations, other methods may have an advantage.
Our discussion will allow the reader to make an informed choice for the most
appropriate method.

Our implementation of Krotov's method together with the examples and
explanations in this paper, and the pseudocode in Appendix~\ref{apx:pseudocode}
may serve as a reference when implementing Krotov's method in other systems or
languages.
We hope that this will motivate wider adoption of Krotov's method, and the use
of optimal quantum control in general.
As quantum technology matures, optimal control for solving the inherently
difficult design problems will only gain in importance.
Thus, the creation of a high quality open source software stack around optimal
control is paramount.
The \texttt{krotov} package is a contribution to this endeavor.

\paragraph{Funding information and acknowledgements}

M.H.G was supported by the Army Research Laboratory under Cooperative Agreement
Number W911NF-17-2-0147.
The Kassel team gratefully acknowledges financial support from the
Volkswagenstiftung, the European Union's Horizon 2020 research and innovation
programme under the Marie Sklodowska-Curie grant agreement Nr.
765267, and the State Hessen Initiative for the Development of Scientific and
Economic Excellence (LOEWE) within the focus project SMolBits.
We thank Steffen Glaser, Shai Machnes, and Nathan Shammah for fruitful
discussions and comments.

\appendix

\section{The Krotov update equation}%
\label{apx:krotov_update}

The core of Krotov's method is the numerical evaluation of the field update in
each iteration, $\Delta \epsilon^{(i)}_l(t)$ in Eq.~\eqref{eq:eps_updated}.
In the following, we specify $\Delta \epsilon^{(i)}_l(t)$ and discuss how its
discretization leads to a numerical scheme.

\subsection{First order update}%
\label{apx:krotov_update_first_order}

Krotov's method is based on a rigorous examination of the conditions for
calculating the updated fields $\{\epsilon_l^{(i)}(t)\}$ such that
\(J(\{\ket{\phi_k^{(i)}(t)}\}, \{\epsilon_l^{(i)}(t)\}) \leq
J(\{\ket{\phi_k^{(i-1)}(t)}\}, \{\epsilon_l^{(i-1)}(t)\})\) is true
\emph{by construction}~\cite{Krotov.book,KonnovARC99,PalaoPRA2003,SklarzPRA2002,ReichJCP12}.
For a general functional of the form in Eq.~\eqref{eq:functional}, with a convex
final-time functional $J_T$, the condition for monotonic convergence is
\begin{equation}%
  \label{eq:krotov_first_order_proto_update}
  \frac{\partial g_a}{\partial \epsilon_l(t)}\bigg\vert_{(i)}
  =  2 \Im \left[\,
      \sum_{k=1}^{N} \Bigg\langle \chi_k^{(i-1)}(t) \Bigg\vert \Bigg(
      \frac{\partial \Op{H}}{\partial \epsilon_l(t)}\bigg\vert_{
      (i)} \Bigg)
      \Bigg\vert \phi_k^{(i)}(t) \Bigg\rangle
    \right]\,,
\end{equation}
see Ref.~\cite{PalaoPRA2003}.
The notation for the derivative on the right hand side being evaluated at
${(i)}$ should be understood to apply when the control Hamiltonian is not linear
so that $\frac{\partial \Op{H}}{\partial \epsilon_l(t)}$ is still
time-dependent; the derivative must then be evaluated for $\epsilon^{(i)}_l(t)$
-- or, numerically, for $\epsilon^{(i-1)}_l(t) \approx \epsilon^{(i)}_l(t)$.
If there are multiple controls, Eq.~\eqref{eq:krotov_first_order_proto_update}
holds for every control field $\epsilon_l(t)$ independently.

For \(g_a\) as in Eq.~\eqref{eq:g_a}, this results in an
\emph{update} equation~\cite{Tannor92,PalaoPRA2003,SklarzPRA2002},
\begin{equation}%
  \label{eq:krotov_first_order_update}
  \Delta\epsilon^{(i)}_l(t)
  = \frac{S_l(t)}{\lambda_{a,l}} \Im \left[\,
      \sum_{k=1}^{N} \Bigg\langle \chi_k^{(i-1)}(t) \Bigg\vert \Bigg(
      \frac{\partial \Op{H}}{\partial \epsilon_l(t)}
      \bigg\vert_{(i)} \Bigg)
      \Bigg\vert \phi_k^{(i)}(t) \Bigg\rangle
    \right]\,,
\end{equation}
cf.~Eq.~\eqref{eq:sequential_update}, with the equation of motion for the
forward propagation of $\ket{\phi_k^{(i)}}$ under the optimized controls
$\{\epsilon_l^{(i)}(t)\}$ of the iteration $(i)$,
\begin{equation}%
  \label{eq:fw_eqm}
  \frac{\partial}{\partial t} \Ket{\phi_k^{(i)}(t)} =
  -\frac{\mathrm{i}}{\hbar} \Op{H}^{(i)} \Ket{\phi_k^{(i)}(t)}\,.
\end{equation}
The co-states \(\ket{\chi_k^{(i-1)}(t)}\) are propagated backwards in time under
the guess controls of iteration $(i)$, i.e., the optimized controls from the
previous iteration $(i-1)$, as
\begin{equation}%
  \label{eq:bw_eqm}
  \frac{\partial}{\partial t} \Ket{\chi_k^{(i-1)}(t)}
  = -\frac{\mathrm{i}}{\hbar} \Op{H}^{\dagger\,(i-1)} \Ket{\chi_k^{(i-1)}(t)}
    + \left.\frac{\partial g_b}{\partial \Bra{\phi_k}}\right\vert_{(i-1)}\,,
\end{equation}
with the boundary condition of Eq.~\eqref{eq:chi_boundary}.

The coupled equations~(\ref{eq:krotov_first_order_update}--\ref{eq:bw_eqm}) can
be generalized to open system dynamics by replacing Hilbert space states with
density matrices, \(\Op{H}\) with \(\mathrm{i} \Liouville\), and brakets with
Hilbert-Schmidt products, \(\langle  \cdot \vert \cdot \rangle \rightarrow
\langle\!\langle \cdot  \vert \cdot \rangle\!\rangle\). In full generality,
$\Op{H}$ in Eq.~\eqref{eq:krotov_first_order_update} is the operator $H$ on the
right-hand side of whatever the equation of motion for the forward propagation
of the states is, written in the form $\mathrm{i} \hbar \dot\phi = H \phi$,
cf.~Eq.~\eqref{eq:fw_eqm}.
Note also that the backward propagation Eq.~\eqref{eq:bw_eqm} uses the adjoint
$H$, which is relevant both for a dissipative
Liouvillian~\cite{BartanaJCP93,OhtsukiJCP99,GoerzNJP2014} and a non-Hermitian
Hamiltonian~\cite{MullerQIP11,GoerzQST2018}.

\subsection{Second order update}%
\label{apx:krotov_update_second_order}

The update Eq.~\eqref{eq:krotov_first_order_update} assumes that the equation of
motion is linear (\(\Op{H}\) does not depend on the states \(\ket{\phi_k(t)}\)),
the functional \(J_T\) is convex, and no state-dependent constraints are used
(\(g_b\equiv 0\)). When any of these conditions are not fulfilled, it is still
possible to derive an optimization algorithm with monotonic convergence via a
``second order'' term in
Eqs.~(\ref{eq:krotov_first_order_proto_update},~\ref{eq:krotov_first_order_update})~\cite{KonnovARC99,ReichJCP12},
The full update equation then reads
\begin{equation}%
  \label{eq:krotov_second_order_update}
  \begin{split}
  \Delta\epsilon^{(i)}_l(t)
    &= \frac{S_l(t)}{\lambda_{a,l}} \Im \left[\,
        \sum_{k=1}^{N} \Bigg\langle \chi_k^{(i-1)}(t) \Bigg\vert \Bigg(
        \frac{\partial \Op{H}}{\partial \epsilon_l(t)}
        \bigg\vert_{(i)} \Bigg)
        \Bigg\vert \phi_k^{(i)}(t) \Bigg\rangle
      \right. \\ & \qquad \qquad \quad \left.
        + \frac{1}{2} \sigma(t)
        \Bigg\langle \Delta\phi_k^{(i)}(t) \Bigg\vert \Bigg(
        \frac{\partial \Op{H}}{\partial \epsilon_l(t)}
        \bigg\vert_{(i)} \Bigg)
        \Bigg\vert \phi_k^{(i)}(t) \Bigg\rangle
      \right]\,,
  \end{split}
\end{equation}
with
\begin{equation}
  \ket{\Delta \phi_k^{(i)}(t)}
  \equiv \ket{\phi_k^{(i)}(t)} - \ket{\phi_k^{(i-1)}(t)}\,,
\end{equation}
see Ref.~\cite{ReichJCP12} for the full construction of the second-order
condition.

In Eq.~\eqref{eq:krotov_second_order_update}, \(\sigma(t)\) is a scalar function
that must be properly chosen to ensure monotonic convergence.
As shown in Ref.~\cite{ReichJCP12}, it is possible to numerically approximate
\(\sigma(t)\), see Section~\ref{sec:task_second_order} for an example.

\subsection{Time discretization}%
\label{apx:krotov_update_discretization}

\begin{figure}[tb]
\centering \includegraphics{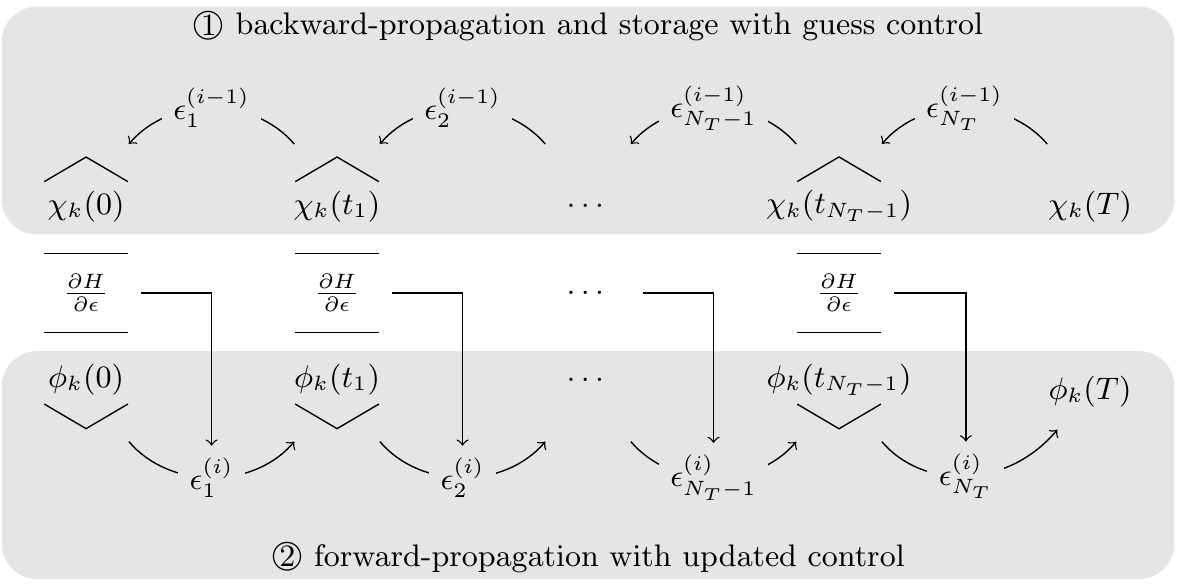}
  \caption{Sequential update scheme in Krotov's method on a time grid.}%
\label{fig:krotovscheme}
\end{figure}

The derivation of Krotov's method assumes time-continuous control fields.
Only in this case, monotonic convergence is mathematically guaranteed.
However, for practical numerical applications, we have to consider controls on
a discrete time grid with \(N_T + 1\) points running from \(t=t_0=0\) to
\(t=t_{N_T} = T\), with a time step \(\dd t\).
The states are defined on the points of the time grid,
while the controls are assumed to be constant on the intervals of the time grid.
A coarse time step must be compensated by larger values of the inverse step size
\(\lambda_{a,l}\), slowing down convergence.
Values that are too small will cause sharp spikes in the optimized control and
numerical instabilities.
A lower limit for \(\lambda_{a,l}\) can be determined from the requirement that
the change \(\Delta\epsilon^{(i)}(t)\) should be at most of the same order of
magnitude as the guess pulse \(\epsilon^{(i-1)}(t)\) for that iteration.
The Cauchy-Schwarz inequality applied to the update
equation~\eqref{eq:krotov_first_order_update} yields
\begin{equation}
  \lambda_a \ge
  \frac{1}{\Norm{\epsilon^{(i-1)}(t)}_{\infty}} \left[ \sum_{k}
    \Norm{\ket{\chi^{(i-1)}_k(t)}}_{\infty} \right]
    \Norm{\frac{\partial \Op{H}}{\partial \epsilon(t)}}_{\infty}\,.
\end{equation}
From a practical point of view, the best strategy is to start the optimization
with a comparatively large value of \(\lambda_{a,l}\), and after a few
iterations lower \(\lambda_{a,l}\) as far as possible without introducing
numerical instabilities.
The value of \(\lambda_{a,l}\) may be adjusted dynamically with respect to the
rate of convergence, via the \texttt{modify\_params\_after\_iter} argument to
the \texttt{optimize\_pulses} function.
Generally, the ideal choice of \(\lambda_{a,l}\) requires some trial and error.

The discretization yields the numerical scheme shown in
Fig.~\ref{fig:krotovscheme} for a single control field (no index $l$), and
assuming the first-order update is sufficient to guarantee monotonic convergence
for the chosen functional.
For simplicity, we also assume that the Hamiltonian is linear in the control, so
that $\partial \Op{H} / \partial \epsilon(t)$ is not time-dependent.
The scheme proceeds as follows:
\begin{enumerate}
  \item
    Construct the states $\{\ket{\chi^{(i-1)}_k(T)}\}$ according to
    Eq.~\eqref{eq:chi_boundary}.
    For most functionals, specifically any that are more than linear in the
    overlaps $\tau_k$ defined in Eq.~\eqref{eq:JTre}, the states
    $\{\ket{\chi^{(i-1)}_k(T)}\}$ depend on the states
    $\{\ket{\phi^{(i-1)}_k(T)}\}$ forward-propagated under the optimized pulse
    from the previous iteration, that is, the guess pulse in the current
    iteration.
  \item
    Perform a backward propagation using Eq.~\eqref{eq:bw_eqm} as the equation
    of motion over the entire time grid.
    The resulting state at each point in the time grid must be stored in memory.
  \item
    Starting from the known initial states $\{\ket{\phi_k}\} =
    \{\ket{\phi_k(t=t_0=0)}\}$, calculate the pulse update for the first time
    step according to
    \begin{equation}%
      \label{eq:update_discretized0}
      \Delta\epsilon^{(i)}_1
      \equiv \Delta\epsilon^{(i)}(\tilde{t}_0)
      = \frac{S(\tilde{t}_0)}{\lambda_{a}} \Im \left[\,
          \sum_{k=1}^{N} \bigg\langle \chi_k^{(i-1)}(t_0) \bigg\vert
          \frac{\partial \Op{H}}{\partial \epsilon}
          \bigg\vert \phi_k(t_0) \bigg\rangle
        \right]\,.
    \end{equation}
    The value $\Delta\epsilon^{(i)}_1 $
    is taken on the midpoint of the first time interval,
    $\tilde{t}_0 \equiv t_0 + \dd t/2$, based on the assumption
    of a piecewise-constant control field and an equidistant time grid with
    spacing $\dd t$.
  \item
    Use the updated field $\epsilon^{(i)}_1$ for the first interval
    to propagate \(\ket{\phi_k(t=t_0)}\) for a single time step to
    \(\ket{\phi_k^{(i)}(t=t_0 + \dd t)}\), with Eq.~\eqref{eq:fw_eqm} as the
    equation of motion.
    The updates then proceed sequentially, using the discretized update
    equation
    \begin{equation}%
      \label{eq:update_discretized}
      \Delta\epsilon^{(i)}_{n+1}
      \equiv \Delta\epsilon^{(i)}(\tilde{t}_n)
      = \frac{S(\tilde{t}_n)}{\lambda_{a}} \Im \left[\,
          \sum_{k=1}^{N} \bigg\langle \chi_k^{(i-1)}(t_n) \bigg\vert
          \frac{\partial \Op{H}}{\partial \epsilon}
          \bigg\vert \phi_k^{(i)}(t_n) \bigg\rangle
        \right]
    \end{equation}
    with $\tilde{t}_n \equiv t_n + \dd t / 2$
    for each time interval $n$, until the final forward-propagated
    state \(\ket{\phi^{(i)}_k(T)}\) is reached.
  \item
    The updated control field becomes the guess control for the next iteration
    of the algorithm, starting again at step 1.
    The optimization continues until the value of the functional $J_T$ falls
    below some predefined threshold, or convergence is reached, i.e., $\Delta
    J_T$ approaches zero so that no further significant improvement of $J_T$ is
    to be expected.
\end{enumerate}
Eq.~\eqref{eq:krotov_first_order_update} re-emerges as the continuous limit of
the time-discretized update equation~\eqref{eq:update_discretized}, i.e.,
$\dd t \rightarrow 0$ so that $\tilde{t}_n \rightarrow t_n$.
Note that Eq.~\eqref{eq:update_discretized} resolves the seeming contradiction
in the time-continuous Eq.~\eqref{eq:krotov_first_order_update} that the
calculation of $\epsilon^{(i)}(t)$ requires knowledge of the states
$\ket{\phi_k^{(i)}(t)}$ which would have to be obtained from a propagation under
$\epsilon^{(i)}(t)$.
By having the time argument $\tilde{t}_n$ on the left-hand-side of
Eq.~\eqref{eq:update_discretized}, and $t_n < \tilde{t}_n$ on the
right-hand-side (with $S(\tilde{t}_n)$ known at all times), the update for each
interval only depends on ``past'' information.

\section{Pseudocode for Krotov's method}%
\label{apx:pseudocode}

\begin{algorithm}
\caption{{\sc Krotov's Method for Quantum Optimal Control}
\label{al:KrotovsMethod} }
  \begin{algorithmic}[1]
    \Require{%
      ~\par
      \begin{enumerate}
      \vspace{-6pt}
      \itemsep0em
      \item
      list of guess control values $\{\VarScalar{\epsilon^{(0)}_{ln}}\}$ where
      $\VarScalar{\epsilon^{(0)}_{ln}}$ is the value of the $l$'th control field
      on the $n$'th interval of the propagation time grid ($t_0 = 0, \dots,
      t_{N_T} = T$),
      i.e., $\VarScalar{\epsilon^{(0)}_{ln}} \equiv
      \epsilon^{(0)}_l(\tilde{t}_{n-1})$ with
      $n \in [1, N_T]$ and $\tilde{t}_n \equiv (t_{n} + t_{n+1})/2$
      \item
      list of update-shape values $\{\VarScalar{S_{ln}}\}$ with each
      $\VarScalar{S_{ln}} \in [0,1]$
      \item
      list of update step size values $\{\VarScalar{\lambda_{a,l}}\}$
      \item
      list of $N$ initial states $\{\VarState{\phi_{k}^{\text{init}}}\}$ at
      $t=t_0=0$
      \item
      propagator function $U$ that in ``forward mode'' receives a state
      $\VarPropState{\phi}{k}{}{t_n}$ and a list of control values
      $\{\VarScalar{\epsilon_{ln}}\}$ and
      returns $\VarPropState{\phi}{k}{}{t_{n+1}}$ by solving the differential
      equation~\eqref{eq:fw_eqm}, respectively in ``backward mode'' (indicated
      as $U^\dagger$) receives a state $\VarPropState{\chi}{k}{}{t_{n}}$ and returns
      $\VarPropState{\chi}{k}{}{t_{n-1}}$ by solving the differential
      equation~\eqref{eq:bw_eqm}
      \item
      list of operators
      $\VarOperator{\mu}{lkn}
      = \frac{\partial H_k}{\partial \epsilon_{l}(t)}\big\vert
        _{\epsilon_{ln}}$,
      cf.~Eq.~\eqref{eq:krotov_first_order_proto_update},
      where $H_k$ is the right-hand-side of the
      equation of motion of $\VarPropState{\phi}{k}{}{t}$, up to a factor of
      $(-\mathrm{i}/\hbar)$, cf.~Eq.~\eqref{eq:fw_eqm}
      \item
      function $\chi$ that receives a list of states $\{\VarPropState{\phi}{k}{}{T}\}$ and
      returns a list of states $\{\VarPropState{\chi}{k}{}{T}\}$ according to
      Eq.~\eqref{eq:chi_boundary}
      \item
      optionally, if  a second order construction of the
      pulse update is necessary: function $\sigma(t)$
      \end{enumerate}
    }
    \vspace{-6pt}
    \Ensure{%
      optimized control values $\{\VarScalar{\epsilon^{(\text{opt})}_{ln}}\}$,
      such that $J[\{\VarScalar{\epsilon_{ln}^{(\text{opt})}}\}]
      \le J[\{\VarScalar{\epsilon_{ln}^{(0)}}\}]$,
      with $J$ defined in Eq.~\eqref{eq:functional}.
    }
    \Procedure{KrotovOptimization}{%
      $\{\VarScalar{\epsilon^{(0)}_{ln}}\}$,
      $\{\VarScalar{S_{ln}}\}$,
      $\{\VarScalar{\lambda_{a,l}}\}$,
      $\{\VarState{\phi_k^{\text{init}}}\}$,
      $U$,
      $\{\VarOperator{\mu}{lkn}\}$,
      $\chi$,
      $\sigma$
    }
      \State $\VarScalar{i} \gets 0$ \Comment{iteration number}
      \State allocate forward storage array $\VarArray{\Phi_0}[1\dots N, 0\dots N_T]$
      \For{$\VarScalar{k} \gets 1, \dots, N$} \Comment{initial forward-propagation}
         \State $\VarArray{\Phi_0}[\VarScalar{k},0] \gets \VarPropState{\phi}{k}{(0)}{t_0} \gets \VarState{\phi_k^{\text{init}}}$
         \For{$\VarScalar{n} \gets 1, \dots, N_T$}
           \State $\VarArray{\Phi_0}[\VarScalar{k}, \VarScalar{n}] \gets \VarPropState{\phi}{k}{(0)}{t_n} \gets U(\VarPropState{\phi}{k}{(0)}{t_{n-1}}, \{\VarScalar{\epsilon^{(0)}_{ln}}\})$ \Comment{propagate and store} \label{ln:prop1}
         \EndFor
      \EndFor
      \While{not converged} \Comment{optimization loop}
         \State $\VarScalar{i} \gets \VarScalar{i} + 1$
         \State $\VarArray{\Phi_1}, \{\VarScalar{\epsilon_{ln}^{(i)}}\} \gets \text{\textsc{KrotovIteration}}(\VarArray{\Phi_0}, \{\VarScalar{\epsilon^{(i-1)}_{ln}}\}, \dots)$
         \State $\VarArray{\Phi_0} \gets \VarArray{\Phi_1}$
      \EndWhile
      \State $\Forall \VarScalar{l}, \Forall \VarScalar{n}: \VarScalar{\epsilon_{ln}^{(\text{opt})}} \gets  \VarScalar{\epsilon_{ln}^{(i)}}$ \Comment{final optimized controls}
    \EndProcedure
    \algstore{krotovalg}
  \end{algorithmic}
\vspace{5pt} \textbf{Notes:}
  \begin{itemize}
  \itemsep0em
    \item The index $\VarScalar{k}$ numbers the independent states to be
      propagated, respectively the independent ``objectives'' (see text for
      details), $\VarScalar{l}$ numbers the independent control fields, and
      $\VarScalar{n}$ numbers the intervals on the time grid.
      All of these indices start at 1.
    \item The optimization loop may be stopped if the optimization functional or
      the change of functional falls below a pre-defined threshold, a maximum
      number of iterations is reached, or any other criterion.
  \end{itemize}
\end{algorithm}

\begin{algorithm}
  \begin{algorithmic}[1]
   \algrestore{krotovalg}
   \Procedure{KrotovIteration}{%
     $\VarArray{\Phi_0}$,
     $\{\VarScalar{\epsilon^{(i-1)}_{ln}}\}$,
     $\{\VarScalar{S_{ln}}\}$,
     $\{\VarScalar{\lambda_{a,l}}\}$,
     $\{\VarState{\phi_k^{\text{init}}}\}$,
     $U$,
     $\{\VarOperator{\mu}{lkn}\}$,
     $\chi$,
     $\sigma$
   }
      \State $\Forall \VarScalar{k}: \VarPropState{\phi}{k}{(i-1)}{T} \gets \VarArray{\Phi_0}[\VarScalar{k}, N_T]$
      \State $\{\VarPropState{\chi}{k}{(i-1)}{T}\} \gets \chi(\{\VarPropState{\phi}{k}{(i-1)}{T}\})$ \Comment{backward boundary condition} \label{ln:chi_boundary}
      \State allocate backward storage array $\VarArray{X}[1 \dots N, 0 \dots N_T]$.
      \For{$\VarScalar{k} \gets 1, \dots, N$}
        \State $\VarArray{X}[\VarScalar{k}, N_T] \gets \VarPropState{\chi}{k}{(i-1)}{T}$
         \For{$\VarScalar{n} \gets N_T, \dots, 1$} \Comment{backward-propagate and store}
           \State $\VarArray{X}[\VarScalar{k}, \VarScalar{n}-1] \gets \VarPropState{\chi}{k}{(i-1)}{t_{n-1}} \gets U^{\dagger}(\VarPropState{\chi}{k}{(i-1)}{t_{n}}, \{\VarScalar{\epsilon^{(i-1)}_{ln}}\}, \VarArray{\Phi_0})$  \label{ln:prop2}
         \EndFor
      \EndFor
      \State allocate forward storage array $\VarArray{\Phi_1}[1\dots N, 0\dots N_T]$
      \State $\Forall \VarScalar{k}: \VarArray{\Phi_1}[\VarScalar{k},0] \gets \VarPropState{\phi}{k}{(i)}{t_0} \gets \VarState{\phi_k^{\text{init}}}$
      \For{$\VarScalar{n} \gets 1, \dots, N_T$} \Comment{sequential update loop}
         \State $\Forall \VarScalar{k}: \VarPropState{\chi}{k}{(i-1)}{t_{n-1}} \gets \VarArray{X}[\VarScalar{k}, \VarScalar{n}-1]$
         \State $\Forall \VarScalar{l}: \VarScalar{\Delta \epsilon_{ln}} \gets \frac{\VarScalar{S_{ln}}}{\VarScalar{\lambda_{a, l}}} \Im \sum_{\VarScalar{k}} \big\langle \VarPropState{\chi}{k}{(i-1)}{t_{n-1}} \big\vert \VarOperator{\mu^{\PropAnnotation{(i-1)}}}{lkn} \big\vert \VarPropState{\phi}{k}{(i)}{t_{n-1}} \big\rangle$ \Comment{first order} \label{ln:pulse_update}
         \If{$\sigma(t) \neq 0$} \Comment{second order}
           \State $\Forall \VarScalar{k}: \VarPropState{\Delta\phi}{k}{(i)}{t_{n-1}} \gets \VarPropState{\phi}{k}{(i)}{t_{n-1}} - \VarArray{\Phi_0}[\VarScalar{k},\VarScalar{n}-1]$ \label{ln:delta_phi}
           \State $\Forall \VarScalar{l}: \VarScalar{\Delta \epsilon_{ln}} \gets \VarScalar{\Delta \epsilon_{ln}} + \frac{\VarScalar{S_{ln}}}{\VarScalar{\lambda_{a, l}}} \Im \sum_{\VarScalar{k}} \frac{1}{2} \sigma(\tilde{t}_n) \big\langle \VarPropState{\Delta\phi}{k}{(i)}{t_{n-1}} \big\vert \VarOperator{\mu^{\PropAnnotation{(i-1)}}}{lkn} \big\vert \VarPropState{\phi}{k}{(i)}{t_{n-1}} \big\rangle$ \label{ln:pulse_update2}
         \EndIf
         \State $\Forall \VarScalar{l}: \VarScalar{\epsilon_{ln}^{(i)}} \gets \VarScalar{\epsilon_{ln}^{(i-1)}} + \VarScalar{\Delta \epsilon_{ln}}$ \Comment{apply update}
         \State $\Forall \VarScalar{k}: \VarArray{\Phi_1}[\VarScalar{k},\VarScalar{n}] \gets \VarPropState{\phi}{k}{(i)}{t_n} \gets U(\VarPropState{\phi}{k}{(i)}{t_{n-1}}, \{\VarScalar{\epsilon_{ln}^{(i)}}\})$ \Comment{propagate and store} \label{ln:prop3}
      \EndFor
      \If{$\sigma(t) \neq 0$}
        \State Update internal parameters of $\sigma(t)$ \label{ln:sigma_update} if necessary, using $\VarArray{\Phi_0}$, $\VarArray{\Phi_1}$
      \EndIf
    \EndProcedure
  \end{algorithmic}
\vspace{5pt} \textbf{Notes:}
  \begin{itemize}
  \itemsep0em
    \item The braket notation in line~\ref{ln:pulse_update} indicates the
      (Hilbert-Schmidt) inner product of the state
      $\VarPropState{\chi}{k}{(i-1)}{t_n-1}$ and the state
      resulting from applying $\VarOperator{\mu^{\PropAnnotation{(i-1)}}}{lkn}$
      to $\VarPropState{\phi}{k}{(i)}{t_{n-1}}$.
      In Hilbert space, this is the standard braket.
      In Liouville space, it is $\tr\left(\VarState{\chi_k}^\dagger
      \;\VarOperator{\mu}{lkn}[\VarState{\phi_k}]\right)$ with density
      matrices $\VarState{\chi_k}$, $\VarState{\phi_k}$ and a super-operator
      $\VarOperator{\mu}{lkn}$.
    \item For numerical stability, the states $\VarPropState{\chi}{k}{(i-1)}{T}$
      in line~\ref{ln:chi_boundary} may be normalized. This norm then has to
      taken into account in the pulse update, line~\ref{ln:pulse_update}.
    \item In line~\ref{ln:prop2}, the storage array $\VarArray{\Phi_0}$ is
      passed to $U^\dagger$ only to account for the inhomogeneity due to
      a possible state-dependent constraint, $\partial g_b / \partial
      \bra{\phi_k}$ in Eq.~\eqref{eq:bw_eqm}.
      If $g_b \equiv 0$, the parameter can be omitted.
  \end{itemize}
\end{algorithm}

For reference, Algorithm~\ref{al:KrotovsMethod} shows the complete pseudocode of
an optimization with Krotov's method, as implemented in the \texttt{krotov}
package.
It realizes the time-discretized scheme described in
Appendix~\ref{apx:krotov_update_discretization}.

Variables are color coded.
Scalars are set in blue, e.g.\ $\VarScalar{\epsilon_{ln}^{(0)}}$.
States  (Hilbert space states or vectorized density matrices) are set in purple,
e.g.\ $\VarState{\phi_k^{\text{init}}}$.
They may be annotated with light gray superscripts to indicate the
iteration-index $\VarScalar{i}$ of the control under which state was propagated,
and with light gray time arguments.
These annotations serve only to connect the variables to the equations in
Appendix~\ref{apx:krotov_update}: $\VarPropState{\phi}{k}{(0)}{t_n}$ and
$\VarPropState{\phi}{k}{(0)}{t_{n-1}}$ are the same variable
$\VarState{\phi_k}$.
Operators acting on states are set in green, e.g.\ $\VarOperator{\mu}{lkn}$.
These may be implemented as a sparse matrix or implicitly as a function that
returns the result of applying the operator to a state.
Lastly, storage arrays are set in red, e.g.\ $\VarArray{\Phi_0}$.
Each element of a storage array is a state.

The Python implementation groups several of the algorithm's input parameters by
introducing a list of $N$ ``objectives''.
The objectives are indexed by $\VarScalar{k}$, and each objective contains the
initial state $\VarState{\phi_k^{\text{init}}}$, the Hamiltonian or Liouvillian
$H_k$ to be used by the propagator $U$ and for the operators
$\VarOperator{\mu}{lkn}$, and possibly a ``target'' to be taken into account by
the function $\chi$.
In many applications, $H_k \equiv H$ is the same in all objectives, and
$\VarOperator{\mu}{lkn} \equiv \VarOperator{\mu}{l}$ if $H$ is linear in the
controls in addition.
The subscript $n$ and the superscript $(i-1)$ for
$\VarOperator{\mu^{\PropAnnotation{(i-1)}}}{lkn}$ in
lines~\ref{ln:pulse_update},~\ref{ln:pulse_update2} comes into play only if $H$
is \emph{not} linear in the control.
Mathematically, $\VarOperator{\mu}{lkn}$ would then have to be evaluated using
the \emph{updated} control.
Since the update is not yet known, the \emph{guess} control may be used as an
approximation (valid for sufficiently large $\lambda_{a,l}$).

The CPU resources required for the optimization are dominated by the time
propagation (calls to the function $U$ in
lines~\ref{ln:prop1},~\ref{ln:prop2}~\ref{ln:prop3}). This is under the
assumption that evaluating $U$ dominates the application of the operator
$\VarOperator{\mu^{\PropAnnotation{(i-1)}}}{lkn}$
to the state $\VarPropState{\phi}{k}{(i)}{t_{n-1}}$ and
the evaluation of the inner product of two states,
lines~\ref{ln:pulse_update},~\ref{ln:pulse_update2}.
This condition is fulfilled for any non-trivial Hilbert space dimension.

Loops over the index $\VarScalar{k}$ are parallelizable, in particular in a
shared-memory (multi-threaded) parallelization environment like OpenMP. In a
(multi-process) method-passing environment like MPI, some care must be taken to
minimize communication overhead from passing large state vectors.
For some (but not all) functionals, inter-process communication can be reduced
to only the scalar values constituting the sum over $\VarScalar{k}$ in
lines~\ref{ln:pulse_update},~\ref{ln:pulse_update2}.

The memory requirements of the algorithm are dominated by the storage arrays
$\VarArray{\Phi_0}$, $\VarArray{\Phi_1}$, and $\VarArray{X}$.
Each of these must store $N (N_T + 1)$ full state vectors (a full time
propagation for each of the $N$ objectives). Each state vector is typically an
array of double-precision complex numbers.
For a Hilbert space dimension $d$, a state vector thus requires $16 d$ bytes of
memory, or $16 d^2$ bytes for a density matrix.
Under certain conditions, the use of $\VarArray{\Phi_0}$ and $\VarArray{\Phi_1}$
can be avoided: both are required only when the second order update is used
($\sigma(t) \neq 0$). When the first order update is sufficient,
$\VarArray{\Phi_1}$ may overwrite $\VarArray{\Phi_0}$ so that the two collapse
into a single forward-storage $\VarArray{\Phi}$.
The states stored in $\VarArray{\Phi}$ are only used for the inhomogeneity
$\partial g_b / \partial \bra{\phi_k}$ in Eq.~\eqref{eq:bw_eqm}, and no storage
$\VarArray{\Phi}$ of forward-propagated states at all is required if $g_b \equiv
0$.
Thus, in most examples, only the storage $\VarArray{X}$ of the
backward-propagated states remains.
In principle, if the time propagation $U$ is unitary (i.e., invertible), the
states stored in $\VarArray{X}$ could be recovered by forward-propagation of
$\{\VarPropState{\chi}{k}{(i-1)}{t=0}\}$, eliminating $\VarArray{X}$ at the
(considerable) runtime cost of an additional time propagation.

\section{Installation instructions}%
\label{apx:installation_instructions}

The \texttt{krotov} package is available for Python versions $\ge$3.5.
Its main dependency is QuTiP~\cite{JohanssonCPC2012, JohanssonCPC2013}.
Thus, you should consider QuTiP's installation instructions, see
\url{http://qutip.org}.

It is strongly recommended to install Python packages into an isolated
environment.
One possible system for creating such environments it \texttt{conda}, available
as part of the Anaconda Python Distribution, respectively the smaller
``Miniconda'', available at \url{https://conda.io/miniconda.html}.
Anaconda has the additional benefit that it provides binary versions of
scientific Python packages that include compiled extensions, and may be hard to
install on systems that lack the necessary compilers (Windows, macOS). This
includes the QuTiP package.
Assuming \texttt{conda} is installed, the following commands set up a virtual
environment into which the {\tt krotov} package can then be installed:
\begin{lstlisting}[language=bash]
$ conda create -n qucontrolenv python=3.7
$ conda activate qucontrolenv
$ conda config --append channels conda-forge
$ conda install qutip
\end{lstlisting}
To install the latest released version of \texttt{krotov} into your current
(conda) environment, run this command in your terminal:
\begin{lstlisting}[language=bash]
$ pip install krotov
\end{lstlisting}
The examples in the online documentation and in
Section~\ref{sec:overview_example} require additional dependencies.
These can be installed with
\begin{lstlisting}[language=bash]
$ pip install krotov[dev]
\end{lstlisting}
See the package documentation linked in Appendix~\ref{apx:package_docs} for the
most current installation instructions.

\section{Package documentation}%
\label{apx:package_docs}

This paper describes only the most central features of the \texttt{krotov}
package.
For a complete documentation, refer to \url{https://qucontrol.github.io/krotov}.
The most current version of the \texttt{krotov} package is available at
\url{https://github.com/qucontrol/krotov} under a BSD license.

The example script of Section~\ref{sec:overview_example} is available at
\url{https://github.com/qucontrol/krotov/tree/paper/examples}.
A Jupyter notebook version of the same example is available in
the Examples section of the online documentation, together with notebooks
illustrating in more detail the optimization tasks discussed in
Section~\ref{sec:common_optimization_tasks}.

\nolinenumbers

\bibliography{refs.bib}

\begin{thebibliography}{100}
\providecommand{\url}[1]{\texttt{#1}}
\providecommand{\urlprefix}{URL }
\expandafter\ifx\csname urlstyle\endcsname\relax
  \providecommand{\doi}[1]{doi:\discretionary{}{}{}#1}\else
  \providecommand{\doi}{doi:\discretionary{}{}{}\begingroup
  \urlstyle{rm}\Url}\fi
\providecommand{\eprint}[2][]{\url{#2}}

\bibitem{AcinNJP18}
A.~Acín, I.~Bloch, H.~Buhrman, T.~Calarco, C.~Eichler, J.~Eisert, D.~Esteve,
  N.~Gisin, S.~J. Glaser, F.~Jelezko, S.~Kuhr, M.~Lewenstein \emph{et~al.},
\newblock \emph{{The quantum technologies roadmap: a European community view}},
\newblock New J. Phys. \textbf{20}, 080201 (2018),
\newblock \doi{10.1088/1367-2630/aad1ea}.

\bibitem{NielsenChuang}
M.~A. Nielsen and I.~L. Chuang,
\newblock \emph{Quantum Computation and Quantum Information},
\newblock Cambridge University Press (2000).

\bibitem{DegenRMP17}
C.~L. Degen, F.~Reinhard and P.~Cappellaro,
\newblock \emph{Quantum sensing},
\newblock Rev. Mod. Phys. \textbf{89}, 035002 (2017),
\newblock \doi{10.1103/RevModPhys.89.035002}.

\bibitem{GlaserEPJD2015}
S.~J. Glaser, U.~Boscain, T.~Calarco, C.~P. Koch, W.~K\"ockenberger,
  R.~Kosloff, I.~Kuprov, B.~Luy, S.~Schirmer, T.~Schulte-Herbr\"uggen, D.~Sugny
  and F.~K. Wilhelm,
\newblock \emph{{Training Schr{\"o}dinger's cat: quantum optimal control}},
\newblock Eur. Phys. J. D \textbf{69}, 279 (2015),
\newblock \doi{10.1140/epjd/e2015-60464-1}.

\bibitem{Tannor92}
D.~Tannor, V.~Kazakov and V.~Orlov,
\newblock \emph{Control of photochemical branching: Novel procedures for
  finding optimal pulses and global upper bounds},
\newblock In J.~Broeckhove and L.~Lathouwers, eds., \emph{Time-dependent
  quantum molecular dynamics}, pp. 347--360. Plenum (1992).

\bibitem{GrossJCP92}
P.~Gross, D.~Neuhauser and H.~Rabitz,
\newblock \emph{Optimal control of curve-crossing systems},
\newblock J. Chem. Phys. \textbf{96}, 2834 (1992),
\newblock \doi{10.1063/1.461980}.

\bibitem{MurdochJMR87}
J.~B. Murdoch, A.~H. Lent and M.~R. Kritzer,
\newblock \emph{Computer-optimized narrowband pulses for multislice imaging},
\newblock J. Magnet. Res. \textbf{74}, 226 (1987),
\newblock \doi{10.1016/0022-2364(87)90336-2}.

\bibitem{GlaserCPL89}
S.~J. Glaser and G.~P. Drobny,
\newblock \emph{{The tailored TOCSY experiment: Chemical shift selective
  coherence transfer}},
\newblock Chem. Phys. Lett. \textbf{164}, 456 (1989),
\newblock \doi{10.1016/0009-2614(89)85238-8}.

\bibitem{KochJPCM16}
C.~P. Koch,
\newblock \emph{Controlling open quantum systems: tools, achievements, and
  limitations},
\newblock J. Phys.: Condens. Matter \textbf{28}, 213001 (2016),
\newblock \doi{10.1088/0953-8984/28/21/213001}.

\bibitem{CuiQST17}
J.~Cui, R.~van Bijnen, T.~Pohl, S.~Montangero and T.~Calarco,
\newblock \emph{Optimal control of {Rydberg} lattice gases},
\newblock Quantum Sci. Technol. \textbf{2}, 035006 (2017),
\newblock \doi{10.1088/2058-9565/aa7daf}.

\bibitem{PatschPRA18}
S.~Patsch, D.~M. Reich, J.-M. Raimond, M.~Brune, S.~Gleyzes and C.~P. Koch,
\newblock \emph{Fast and accurate circularization of a {Rydberg} atom},
\newblock Phys. Rev. A \textbf{97}, 053418 (2018),
\newblock \doi{10.1103/PhysRevA.97.053418}.

\bibitem{LovecchioPRA16}
C.~Lovecchio, F.~Sch\"afer, S.~Cherukattil, M.~Alí~Khan, I.~Herrera, F.~S.
  Cataliotti, T.~Calarco, S.~Montangero and F.~Caruso,
\newblock \emph{Optimal preparation of quantum states on an atom-chip device},
\newblock Phys. Rev. A \textbf{93}, 010304 (2016),
\newblock \doi{10.1103/PhysRevA.93.010304}.

\bibitem{vanFrankSciRep16}
S.~van Frank, M.~Bonneau, J.~Schmiedmayer, S.~Hild, C.~Gross, M.~Cheneau,
  I.~Bloch, T.~Pichler, A.~Negretti, T.~Calarco and S.~Montangero,
\newblock \emph{Optimal control of complex atomic quantum systems},
\newblock Sci. Rep. \textbf{6}, 34187 (2016),
\newblock \doi{10.1038/srep34187}.

\bibitem{OfekNat16}
N.~Ofek, A.~Petrenko, R.~Heeres, P.~Reinhold, Z.~Leghtas, B.~Vlastakis, Y.~Liu,
  L.~Frunzio, S.~M. Girvin, L.~Jiang, M.~Mirrahimi, M.~H. Devoret
  \emph{et~al.},
\newblock \emph{Extending the lifetime of a quantum bit with error correction
  in superconducting circuits},
\newblock Nature \textbf{536}, 441 (2016),
\newblock \doi{10.1038/nature18949}.

\bibitem{SorensenNat16}
J.~J. W.~H. S{\o}rensen, M.~K. Pedersen, M.~Munch, P.~Haikka, J.~H. Jensen,
  T.~Planke, M.~G. Andreasen, M.~Gajdacz, K.~M{\o}lmer, A.~Lieberoth, J.~F.
  Sherson and Q.~M. players,
\newblock \emph{Exploring the quantum speed limit with computer games},
\newblock Nature \textbf{532}, 210 (2016),
\newblock \doi{10.1038/nature17620}.

\bibitem{HeeresNatComm17}
R.~W. Heeres, P.~Reinhold, N.~Ofek, L.~Frunzio, L.~Jiang, , M.~H. Devoret and
  R.~J. Schoelkopf,
\newblock \emph{Implementing a universal gate set on a logical qubit encoded in
  an oscillator},
\newblock Nature Commun. \textbf{8}, 94 (2017),
\newblock \doi{10.1038/s41467-017-00045-1}.

\bibitem{HeckPNAS18}
R.~Heck, O.~Vuculescu, J.~J. S{\o}rensen, J.~Zoller, M.~G. Andreasen, M.~G.
  Bason, P.~Ejlertsen, O.~El{\'\i}asson, P.~Haikka, J.~S. Laustsen, L.~L.
  Nielsen, A.~Mao \emph{et~al.},
\newblock \emph{Remote optimization of an ultracold atoms experiment by experts
  and citizen scientists},
\newblock Proc. Nat. Acad. Sci. \textbf{115}, E11231 (2018),
\newblock \doi{10.1073/pnas.1716869115}.

\bibitem{FengPRA18}
G.~Feng, F.~H. Cho, H.~Katiyar, J.~Li, D.~Lu, J.~Baugh and R.~Laflamme,
\newblock \emph{Gradient-based closed-loop quantum optimal control in a
  solid-state two-qubit system},
\newblock Phys. Rev. A \textbf{98}, 052341 (2018),
\newblock \doi{10.1103/PhysRevA.98.052341}.

\bibitem{OmranS2019}
A.~Omran, H.~Levine, A.~Keesling, G.~Semeghini, T.~T. Wang, S.~Ebadi,
  H.~Bernien, A.~S. Zibrov, H.~Pichler, S.~Choi, J.~Cui, M.~Rossignolo
  \emph{et~al.},
\newblock \emph{Generation and manipulation of {Schr{\"o}dinger} cat states in
  {Rydberg} atom arrays},
\newblock Science \textbf{365}, 570 (2019),
\newblock \doi{10.1126/science.aax9743}.

\bibitem{Larrouy}
A.~Larrouy, S.~Patsch \emph{et~al.},
\newblock in preparation.

\bibitem{KhanejaJMR05}
N.~Khaneja, T.~Reiss, C.~Kehlet, T.~Schulte-Herbr{\"u}ggen and S.~J. Glaser,
\newblock \emph{{Optimal control of coupled spin dynamics: design of NMR pulse
  sequences by gradient ascent algorithms}},
\newblock J. Magnet. Res. \textbf{172}, 296 (2005),
\newblock \doi{10.1016/j.jmr.2004.11.004}.

\bibitem{ReichJCP12}
D.~M. Reich, M.~Ndong and C.~P. Koch,
\newblock \emph{Monotonically convergent optimization in quantum control using
  {Krotov's} method},
\newblock J. Chem. Phys. \textbf{136}, 104103 (2012),
\newblock \doi{10.1063/1.3691827}.

\bibitem{EitanPRA11}
R.~Eitan, M.~Mundt and D.~J. Tannor,
\newblock \emph{Optimal control with accelerated convergence: Combining the
  {Krotov} and quasi-{Newton} methods},
\newblock Phys. Rev. A \textbf{83}, 053426 (2011),
\newblock \doi{10.1103/PhysRevA.83.053426}.

\bibitem{JohanssonCPC2012}
J.~Johansson, P.~Nation and F.~Nori,
\newblock \emph{{QuTiP}: An open-source {Python} framework for the dynamics of
  open quantum systems},
\newblock Comput. Phys. Commun. \textbf{183}, 1760 (2012),
\newblock \doi{10.1016/j.cpc.2012.02.021}.

\bibitem{JohanssonCPC2013}
J.~Johansson, P.~Nation and F.~Nori,
\newblock \emph{{QuTiP 2}: A {Python} framework for the dynamics of open
  quantum systems},
\newblock Comput. Phys. Commun. \textbf{184}, 1234 (2013),
\newblock \doi{10.1016/j.cpc.2012.11.019},
\newblock \urlprefix\url{http://qutip.org}.

\bibitem{CanevaPRA11}
T.~Caneva, T.~Calarco and S.~Montangero,
\newblock \emph{Chopped random-basis quantum optimization},
\newblock Phys. Rev. A \textbf{84}, 022326 (2011),
\newblock \doi{10.1103/PhysRevA.84.022326}.

\bibitem{Jupyter}
T.~Kluyver, B.~Ragan-Kelley, F.~P{\'e}rez, B.~Granger, M.~Bussonnier,
  J.~Frederic, K.~Kelley, J.~Hamrick, J.~Grout, S.~Corlay, P.~Ivanov, D.~Avila
  \emph{et~al.},
\newblock \emph{{Jupyter Notebooks} - a publishing format for reproducible
  computational workflows},
\newblock In F.~Loizides and B.~Schmidt, eds., \emph{Positioning and Power in
  Academic Publishing: Players, Agents and Agendas}, p.~87. IOS Press,
\newblock \doi{10.3233/978-1-61499-649-1-87},
\newblock \urlprefix\url{https://jupyter.org} (2016).

\bibitem{MullerQIP11}
M.~M. M{\"u}ller, H.~Haakh, T.~Calarco, C.~P. Koch and C.~Henkel,
\newblock \emph{Prospects for fast {Rydberg} gates on an atom chip},
\newblock Quantum Inf. Process. \textbf{10}, 771 (2011),
\newblock \doi{10.1007/s11128-011-0296-0}.

\bibitem{GoerzPRA2014}
M.~H. Goerz, E.~J. Halperin, J.~M. Aytac, C.~P. Koch and K.~B. Whaley,
\newblock \emph{Robustness of high-fidelity {Rydberg} gates with single-site
  addressability},
\newblock Phys. Rev. A \textbf{90}, 032329 (2014),
\newblock \doi{10.1103/PhysRevA.90.032329}.

\bibitem{GoerzNJP2014}
M.~H. Goerz, D.~M. Reich and C.~P. Koch,
\newblock \emph{Optimal control theory for a unitary operation under
  dissipative evolution},
\newblock New J. Phys. \textbf{16}, 055012 (2014),
\newblock \doi{10.1088/1367-2630/16/5/055012}.

\bibitem{WattsPRA2015}
P.~Watts, J.~Vala, M.~M. M\"uller, T.~Calarco, K.~B. Whaley, D.~M. Reich, M.~H.
  Goerz and C.~P. Koch,
\newblock \emph{Optimizing for an arbitrary perfect entangler: {I.
  Functionals}},
\newblock Phys. Rev. A \textbf{91}, 062306 (2015),
\newblock \doi{10.1103/PhysRevA.91.062306}.

\bibitem{GoerzPRA2015}
M.~H. Goerz, G.~Gualdi, D.~M. Reich, C.~P. Koch, F.~Motzoi, K.~B. Whaley,
  J.~Vala, M.~M. M\"uller, S.~Montangero and T.~Calarco,
\newblock \emph{Optimizing for an arbitrary perfect entangler. {II.
  Application}},
\newblock Phys. Rev. A \textbf{91}, 062307 (2015),
\newblock \doi{10.1103/PhysRevA.91.062307}.

\bibitem{BasilewitschNJP2017}
D.~Basilewitsch, R.~Schmidt, D.~Sugny, S.~Maniscalco and C.~P. Koch,
\newblock \emph{Beating the limits with initial correlations},
\newblock New J. Phys. \textbf{19}, 113042 (2017),
\newblock \doi{10.1088/1367-2630/aa96f8}.

\bibitem{PreskillQ2018}
J.~Preskill,
\newblock \emph{Quantum computing in the {NISQ} era and beyond},
\newblock Quantum \textbf{2}, 79 (2018),
\newblock \doi{10.22331/q-2018-08-06-79}.

\bibitem{GoerzEPJQT2015}
M.~H. Goerz, K.~B. Whaley and C.~P. Koch,
\newblock \emph{Hybrid optimization schemes for quantum control},
\newblock EPJ Quantum Technol. \textbf{2}, 21 (2015),
\newblock \doi{10.1140/epjqt/s40507-015-0034-0}.

\bibitem{GoerzNPJQI17}
M.~H. Goerz, F.~Motzoi, K.~B. Whaley and C.~P. Koch,
\newblock \emph{Charting the circuit-{QED} design landscape using optimal
  control theory},
\newblock npj Quantum Inf. \textbf{3}, 37 (2017),
\newblock \doi{10.1038/s41534-017-0036-0}.

\bibitem{BezansonSIREV2017}
J.~Bezanson, A.~Edelman, S.~Karpinski and V.~Shah,
\newblock \emph{Julia: A fresh approach to numerical computing},
\newblock SIAM Rev. \textbf{59}, 65 (2017),
\newblock \doi{10.1137/141000671}.

\bibitem{AkeretAC2015}
J.~Akeret, L.~Gamper, A.~Amara and A.~Refregier,
\newblock \emph{{HOPE}: A {Python} just-in-time compiler for astrophysical
  computations},
\newblock Astron. Comput. \textbf{10}, 1 (2015),
\newblock \doi{10.1016/j.ascom.2014.12.001}.

\bibitem{EichhornCSJ2018}
H.~Eichhorn, J.~L. Cano, F.~McLean and R.~Anderl,
\newblock \emph{A comparative study of programming languages for
  next-generation astrodynamics systems},
\newblock CEAS Space J. \textbf{10}, 115 (2018),
\newblock \doi{10.1007/s12567-017-0170-8}.

\bibitem{TannorJCP1985}
D.~J. Tannor and S.~A. Rice,
\newblock \emph{Control of selectivity of chemical reaction via control of wave
  packet evolution},
\newblock J. Chem. Phys. \textbf{83}, 5013 (1985),
\newblock \doi{10.1063/1.449767}.

\bibitem{BellmanBook}
R.~Bellman,
\newblock \emph{Dynamic Programming},
\newblock Princeton University Press, Princeton, NJ (1957).

\bibitem{PontryaginBook}
L.~S. Pontryagin, V.~G. Boltyanskii, G.~R. V. and E.~F. Mishchenko,
\newblock \emph{The Mathematical Theory of Optimal Processes},
\newblock Interscience, New York, NY (1962).

\bibitem{KrotovEC1983}
V.~F. Krotov and I.~N. Fel'dman,
\newblock \emph{An iterative method for solving optimal-control problems},
\newblock Engrg. Cybernetics \textbf{21}, 123 (1983).

\bibitem{KrotovCC1988}
V.~F. Krotov,
\newblock \emph{A technique of global bounds in optimal control theory},
\newblock Control and Cybernetics \textbf{17}, 115 (1988).

\bibitem{Krotov.book}
V.~Krotov,
\newblock \emph{Global Methods in Optimal Control Theory},
\newblock CRC Press (1995).

\bibitem{KonnovARC99}
A.~Konnov and V.~F. Krotov,
\newblock \emph{On global methods of successive improvement of controlled
  processes},
\newblock Autom. Rem. Contr. \textbf{60}, 1427 (1999).

\bibitem{SomloiCP1993}
J.~Somlói, V.~A. Kazakov and D.~J. Tannor,
\newblock \emph{Controlled dissociation of {I2} via optical transitions between
  the {X} and {B} electronic states},
\newblock Chem. Phys. \textbf{172}, 85 (1993),
\newblock \doi{10.1016/0301-0104(93)80108-L}.

\bibitem{BartanaJCP1997}
A.~Bartana, R.~Kosloff and D.~J. Tannor,
\newblock \emph{Laser cooling of internal degrees of freedom. {II}},
\newblock J. Chem. Phys. \textbf{106}, 1435 (1997),
\newblock \doi{10.1063/1.473973}.

\bibitem{SklarzPRA2002}
S.~E. Sklarz and D.~J. Tannor,
\newblock \emph{Loading a {Bose-Einstein} condensate onto an optical lattice:
  An application of optimal control theory to the nonlinear {Schr\"odinger}
  equation},
\newblock Phys. Rev. A \textbf{66}, 053619 (2002),
\newblock \doi{10.1103/PhysRevA.66.053619}.

\bibitem{PalaoPRA2003}
J.~P. Palao and R.~Kosloff,
\newblock \emph{Optimal control theory for unitary transformations},
\newblock Phys. Rev. A \textbf{68}, 062308 (2003),
\newblock \doi{10.1103/PhysRevA.68.062308}.

\bibitem{KaiserJCP2004}
A.~Kaiser and V.~May,
\newblock \emph{Optimal control theory for a target state distributed in time:
  Optimizing the probe-pulse signal of a pump-probe-scheme},
\newblock J. Chem. Phys. \textbf{121}, 2528 (2004),
\newblock \doi{10.1063/1.1769370}.

\bibitem{SerbanPRA2005}
I.~Serban, J.~Werschnik and E.~K.~U. Gross,
\newblock \emph{Optimal control of time-dependent targets},
\newblock Phys. Rev. A \textbf{71}, 053810 (2005),
\newblock \doi{10.1103/PhysRevA.71.053810}.

\bibitem{PalaoPRA2008}
J.~P. Palao, R.~Kosloff and C.~P. Koch,
\newblock \emph{Protecting coherence in optimal control theory: State-dependent
  constraint approach},
\newblock Phys. Rev. A \textbf{77}, 063412 (2008),
\newblock \doi{10.1103/PhysRevA.77.063412}.

\bibitem{van-der-WaltCSE2011}
S.~{van der Walt}, S.~C. {Colbert} and G.~{Varoquaux},
\newblock \emph{The numpy array: A structure for efficient numerical
  computation},
\newblock Comput. Sci. Eng. \textbf{13}, 22 (2011),
\newblock \doi{10.1109/MCSE.2011.37},
\newblock \urlprefix\url{http://www.numpy.org}.

\bibitem{JakschPRL2000}
D.~Jaksch, J.~I. Cirac, P.~Zoller, S.~L. Rolston, R.~C\^ot\'e and M.~D. Lukin,
\newblock \emph{Fast quantum gates for neutral atoms},
\newblock Phys. Rev. Lett. \textbf{85}, 2208 (2000),
\newblock \doi{10.1103/PhysRevLett.85.2208}.

\bibitem{MullerPRA11}
M.~M. M{\"u}ller, D.~M. Reich, M.~Murphy, H.~Yuan, J.~Vala, K.~B. Whaley,
  T.~Calarco and C.~P. Koch,
\newblock \emph{Optimizing entangling quantum gates for physical systems},
\newblock Phys. Rev. A \textbf{84}, 042315 (2011),
\newblock \doi{10.1103/PhysRevA.84.042315}.

\bibitem{WattsE2013}
P.~Watts, M.~O'Connor and J.~Vala,
\newblock \emph{Metric structure of the space of two-qubit gates, perfect
  entanglers and quantum control},
\newblock Entropy \textbf{15}, 1963 (2013),
\newblock \doi{10.3390/e15061963}.

\bibitem{MuszPRA2013}
M.~Musz, M.~Ku\'{s} and K.~\.{Z}yczkowski,
\newblock \emph{Unitary quantum gates, perfect entanglers, and unistochastic
  maps},
\newblock Phys. Rev. A \textbf{87}, 022111 (2013),
\newblock \doi{10.1103/PhysRevA.87.022111}.

\bibitem{KosloffCP1989}
R.~Kosloff, S.~Rice, P.~Gaspard, S.~Tersigni and D.~Tannor,
\newblock \emph{Wavepacket dancing: Achieving chemical selectivity by shaping
  light pulses},
\newblock Chem. Phys. \textbf{139}, 201 (1989),
\newblock \doi{10.1016/0301-0104(89)90012-8}.

\bibitem{ShiJCP1990}
S.~Shi and H.~Rabitz,
\newblock \emph{Quantum mechanical optimal control of physical observables in
  microsystems},
\newblock J. Chem. Phys. \textbf{92}, 364 (1990),
\newblock \doi{10.1063/1.458438}.

\bibitem{ShiCPC1991}
S.~Shi and H.~Rabitz,
\newblock \emph{Optimal control of bond selectivity in unimolecular reactions},
\newblock Comput. Phys. Commun. \textbf{63}, 71 (1991),
\newblock \doi{10.1016/0010-4655(91)90239-H}.

\bibitem{Tannor91}
D.~J. Tannor and Y.~Jin,
\newblock \emph{Design of femtosecond pulse sequences to control photochemical
  products},
\newblock In \emph{Mode Selective Chemistry}, pp. 333--345. Springer (1991).

\bibitem{ZhuJCP98}
W.~Zhu, J.~Botina and H.~Rabitz,
\newblock \emph{Rapidly convergent iteration methods for quantum optimal
  control of population},
\newblock J. Chem. Phys. \textbf{108}, 1953 (1998),
\newblock \doi{10.1063/1.475576}.

\bibitem{MadayJCP2003}
Y.~Maday and G.~Turinici,
\newblock \emph{New formulations of monotonically convergent quantum control
  algorithms},
\newblock J. Chem. Phys. \textbf{118}, 8191 (2003),
\newblock \doi{10.1063/1.1564043}.

\bibitem{OhtsukiJCP2004}
Y.~Ohtsuki, G.~Turinici and H.~Rabitz,
\newblock \emph{Generalized monotonically convergent algorithms for solving
  quantum optimal control problems},
\newblock J. Chem. Phys. \textbf{120}, 5509 (2004),
\newblock \doi{10.1063/1.1650297}.

\bibitem{WerschnikJPB2007}
J.~Werschnik and E.~K.~U. Gross,
\newblock \emph{Quantum optimal control theory},
\newblock J. Phys. B \textbf{40}, R175 (2007),
\newblock \doi{10.1088/0953-4075/40/18/r01}.

\bibitem{SchmidtCPC2018}
B.~Schmidt and C.~Hartmann,
\newblock \emph{{WavePacket}: A {Matlab} package for numerical quantum
  dynamics. {II}: Open quantum systems, optimal control, and model reduction},
\newblock Comput. Phys. Commun. \textbf{228}, 229 (2018),
\newblock \doi{10.1016/j.cpc.2018.02.022}.

\bibitem{SchirmerNJP2011}
S.~G. Schirmer and P.~de~Fouquieres,
\newblock \emph{Efficient algorithms for optimal control of quantum dynamics:
  the krotov method unencumbered},
\newblock New J. Phys. \textbf{13}, 073029 (2011),
\newblock \doi{10.1088/1367-2630/13/7/073029}.

\bibitem{MachnesPRA11}
S.~Machnes, U.~Sander, S.~J. Glaser, P.~de~Fouqui\`eres, A.~Gruslys,
  S.~Schirmer and T.~Schulte-Herbr\"uggen,
\newblock \emph{Comparing, optimizing, and benchmarking quantum-control
  algorithms in a unifying programming framework},
\newblock Phys. Rev. A \textbf{84}, 022305 (2011),
\newblock \doi{10.1103/PhysRevA.84.022305}.

\bibitem{PalaoPRL2002}
J.~P. Palao and R.~Kosloff,
\newblock \emph{Quantum computing by an optimal control algorithm for unitary
  transformations},
\newblock Phys. Rev. Lett. \textbf{89}, 188301 (2002),
\newblock \doi{10.1103/PhysRevLett.89.188301}.

\bibitem{GoerzPhd2015}
M.~Goerz,
\newblock \emph{Optimizing Robust Quantum Gates in Open Quantum Systems},
\newblock Ph.D. thesis, Universit{\"a}t Kassel,
\newblock
  \urlprefix\url{https://kobra.bibliothek.uni-kassel.de/handle/urn:nbn:de:hebis:34-2015052748381}
  (2015).

\bibitem{SchirmerJMO2009}
S.~Schirmer,
\newblock \emph{Implementation of quantum gates via optimal control},
\newblock J. Mod. Opt. \textbf{56}, 831 (2009),
\newblock \doi{10.1080/09500340802344933}.

\bibitem{FloetherNJP12}
F.~F. Floether, P.~de~Fouquières and S.~G. Schirmer,
\newblock \emph{Robust quantum gates for open systems via optimal control:
  Markovian versus non-markovian dynamics},
\newblock New J. Phys. \textbf{14}, 073023 (2012),
\newblock \doi{10.1088/1367-2630/14/7/073023}.

\bibitem{ByrdSJSC1995}
R.~H. Byrd, P.~Lu, J.~Nocedal and C.~Zhu,
\newblock \emph{A limited memory algorithm for bound constrained optimization},
\newblock SIAM J. Sci. Comput. \textbf{16}, 1190 (1995),
\newblock \doi{10.1137/0916069}.

\bibitem{ZhuATMS97}
C.~Zhu, R.~H. Byrd, P.~Lu and J.~Nocedal,
\newblock \emph{{Algorithm 778: L-BFGS-B: Fortran subroutines for large-scale
  bound-constrained optimization}},
\newblock ACM Trans. Math. Softw. \textbf{23}, 550 (1997),
\newblock \doi{10.1145/279232.279236}.

\bibitem{FouquieresJMR2011}
P.~de~Fouquières, S.~Schirmer, S.~Glaser and I.~Kuprov,
\newblock \emph{Second order gradient ascent pulse engineering},
\newblock J. Magnet. Res. \textbf{212}, 412 (2011),
\newblock \doi{10.1016/j.jmr.2011.07.023}.

\bibitem{Scipy}
E.~Jones, T.~Oliphant, P.~Peterson \emph{et~al.},
\newblock \emph{{SciPy}: Open source scientific tools for {Python}},
\newblock \urlprefix\url{http://www.scipy.org/} (2001--).

\bibitem{JaegerPRA14}
G.~J{\"a}ger, D.~M. Reich, M.~H. Goerz, C.~P. Koch and U.~Hohenester,
\newblock \emph{Optimal quantum control of {Bose-Einstein} condensates in
  magnetic microtraps: Comparison of {GRAPE} and {Krotov} optimization
  schemes},
\newblock Phys. Rev. A \textbf{90}, 033628 (2014),
\newblock \doi{10.1103/PhysRevA.90.033628}.

\bibitem{NevesJMR2009}
J.~L. Neves, B.~Heitmann, N.~Khaneja and S.~J. Glaser,
\newblock \emph{Heteronuclear decoupling by optimal tracking},
\newblock J. Magnet. Res. \textbf{201}, 7 (2009),
\newblock \doi{10.1016/j.jmr.2009.07.024}.

\bibitem{NguyenJMR2017}
T.~T. Nguyen and S.~J. Glaser,
\newblock \emph{An optimal control approach to design entire relaxation
  dispersion experiments},
\newblock J. Magnet. Res. \textbf{282}, 142 (2017),
\newblock \doi{10.1016/j.jmr.2017.07.010}.

\bibitem{AnselPRA2017}
Q.~Ansel, M.~Tesch, S.~J. Glaser and D.~Sugny,
\newblock \emph{Optimizing fingerprinting experiments for parameter
  identification: Application to spin systems},
\newblock Phys. Rev. A \textbf{96}, 053419 (2017),
\newblock \doi{10.1103/PhysRevA.96.053419}.

\bibitem{SpindlerJMR2012}
P.~E. Spindler, Y.~Zhang, B.~Endeward, N.~Gershernzon, T.~E. Skinner, S.~J.
  Glaser and T.~F. Prisner,
\newblock \emph{Shaped optimal control pulses for increased excitation
  bandwidth in epr},
\newblock J. Magnet. Res. \textbf{218}, 49 (2012),
\newblock \doi{10.1016/j.jmr.2012.02.013}.

\bibitem{TosnerACIE2018}
Z.~Tošner, R.~Sarkar, J.~Becker-Baldus, C.~Glaubitz, S.~Wegner, F.~Engelke,
  S.~J. Glaser and B.~Reif,
\newblock \emph{Overcoming volume selectivity of dipolar recoupling in
  biological solid-state {NMR} spectroscopy},
\newblock Angew. Chem. Int. Ed. \textbf{57}, 14514 (2018),
\newblock \doi{10.1002/anie.201805002}.

\bibitem{LeungPRA2017}
N.~Leung, M.~Abdelhafez, J.~Koch and D.~Schuster,
\newblock \emph{Speedup for quantum optimal control from automatic
  differentiation based on graphics processing units},
\newblock Phys. Rev. A \textbf{95}, 042318 (2017),
\newblock \doi{10.1103/PhysRevA.95.042318}.

\bibitem{AbdelhafezPRA2019}
M.~Abdelhafez, D.~I. Schuster and J.~Koch,
\newblock \emph{Gradient-based optimal control of open quantum systems using
  quantum trajectories and automatic differentiation},
\newblock Phys. Rev. A \textbf{99}, 052327 (2019),
\newblock \doi{10.1103/PhysRevA.99.052327}.

\bibitem{WinckelIP2008}
G.~v. Winckel and A.~Borzì,
\newblock \emph{Computational techniques for a quantum control problem with
  {H$^1$}-cost},
\newblock Inverse Problems \textbf{24}, 034007 (2008),
\newblock \doi{10.1088/0266-5611/24/3/034007}.

\bibitem{SkinnerJMR2010}
T.~E. Skinner and N.~I. Gershenzon,
\newblock \emph{Optimal control design of pulse shapes as analytic functions},
\newblock J. Magnet. Res. \textbf{204}, 248 (2010),
\newblock \doi{10.1016/j.jmr.2010.03.002}.

\bibitem{MotzoiPRA2011}
F.~Motzoi, J.~M. Gambetta, S.~T. Merkel and F.~K. Wilhelm,
\newblock \emph{Optimal control methods for rapidly time-varying hamiltonians},
\newblock Phys. Rev. A \textbf{84}, 022307 (2011),
\newblock \doi{10.1103/PhysRevA.84.022307}.

\bibitem{LucarelliPRA2018}
D.~Lucarelli,
\newblock \emph{Quantum optimal control via gradient ascent in function space
  and the time-bandwidth quantum speed limit},
\newblock Phys. Rev. A \textbf{97}, 062346 (2018),
\newblock \doi{10.1103/physreva.97.062346}.

\bibitem{SorensenPRA2018}
J.~J. W.~H. S\o{}rensen, M.~O. Aranburu, T.~Heinzel and J.~F. Sherson,
\newblock \emph{Quantum optimal control in a chopped basis: Applications in
  control of {Bose-Einstein} condensates},
\newblock Phys. Rev. A \textbf{98}, 022119 (2018),
\newblock \doi{10.1103/PhysRevA.98.022119}.

\bibitem{SorensenCPC2019}
J.~J. Sørensen, J.~H.~M. Jensen, T.~Heinzel and J.~F. Sherson,
\newblock \emph{{QEngine}: A {C++} library for quantum optimal control of
  ultracold atoms},
\newblock Comput. Phys. Commun. \textbf{243}, 135 (2019),
\newblock \doi{10.1016/j.cpc.2019.04.020}.

\bibitem{MachnesPRL2018}
S.~Machnes, E.~Ass\'emat, D.~Tannor and F.~K. Wilhelm,
\newblock \emph{Tunable, flexible, and efficient optimization of control pulses
  for practical qubits},
\newblock Phys. Rev. Lett. \textbf{120}, 150401 (2018),
\newblock \doi{10.1103/PhysRevLett.120.150401}.

\bibitem{RachPRA2015}
N.~Rach, M.~M. M\"uller, T.~Calarco and S.~Montangero,
\newblock \emph{Dressing the chopped-random-basis optimization: A
  bandwidth-limited access to the trap-free landscape},
\newblock Phys. Rev. A \textbf{92}, 062343 (2015),
\newblock \doi{10.1103/PhysRevA.92.062343}.

\bibitem{GoetzPRA2016}
R.~E. Goetz, A.~Karamatskou, R.~Santra and C.~P. Koch,
\newblock \emph{Quantum optimal control of photoelectron spectra and angular
  distributions},
\newblock Phys. Rev. A \textbf{93}, 013413 (2016),
\newblock \doi{10.1103/PhysRevA.93.013413}.

\bibitem{HornNJP2018}
K.~P. Horn, F.~Reiter, Y.~Lin, D.~Leibfried and C.~P. Koch,
\newblock \emph{Quantum optimal control of the dissipative production of a
  maximally entangled state},
\newblock New J. Phys. \textbf{20}, 123010 (2018),
\newblock \doi{10.1088/1367-2630/aaf360}.

\bibitem{DoriaPRL11}
P.~Doria, T.~Calarco and S.~Montangero,
\newblock \emph{Optimal control technique for many-body quantum dynamics},
\newblock Phys. Rev. Lett. \textbf{106}, 190501 (2011),
\newblock \doi{10.1103/PhysRevLett.106.190501}.

\bibitem{CanevaPRA2011}
T.~Caneva, T.~Calarco and S.~Montangero,
\newblock \emph{Chopped random-basis quantum optimization},
\newblock Phys. Rev. A \textbf{84}, 022326 (2011),
\newblock \doi{10.1103/PhysRevA.84.022326}.

\bibitem{NLOpt}
S.~G. Johnson,
\newblock \emph{The {NLopt} nonlinear-optimization package},
\newblock http://ab-initio.mit.edu/nlopt.

\bibitem{nevergrad}
J.~Rapin and O.~Teytaud,
\newblock \emph{{Nevergrad - A gradient-free optimization platform}},
\newblock \url{https://GitHub.com/FacebookResearch/Nevergrad} (2018).

\bibitem{PetersenIETCTA2010}
I.~Petersen,
\newblock \emph{Quantum control theory and applications: a survey},
\newblock IET Control Theory \& Applications \textbf{4}, 2651 (2010),
\newblock \doi{10.1049/iet-cta.2009.0508}.

\bibitem{GoerzNPJQI2017}
M.~H. Goerz, F.~Motzoi, K.~B. Whaley and C.~P. Koch,
\newblock \emph{Charting the circuit {QED} design landscape using optimal
  control theory},
\newblock npj Quantum Information \textbf{3}, 37 (2017),
\newblock \doi{10.1038/s41534-017-0036-0}.

\bibitem{CanevaPRL09}
T.~Caneva, M.~Murphy, T.~Calarco, R.~Fazio, S.~Montangero, V.~Giovannetti and
  G.~E. Santoro,
\newblock \emph{Optimal control at the quantum speed limit},
\newblock Phys. Rev. Lett. \textbf{103}, 240501 (2009),
\newblock \doi{10.1103/PhysRevLett.103.240501}.

\bibitem{GoerzJPB11}
M.~H. Goerz, T.~Calarco and C.~P. Koch,
\newblock \emph{The quantum speed limit of optimal controlled phasegates for
  trapped neutral atoms},
\newblock J. Phys. B \textbf{44}, 154011 (2011),
\newblock \doi{10.1088/0953-4075/44/15/154011}.

\bibitem{MuellerPRA2011}
M.~M. M\"uller, D.~M. Reich, M.~Murphy, H.~Yuan, J.~Vala, K.~B. Whaley,
  T.~Calarco and C.~P. Koch,
\newblock \emph{Optimizing entangling quantum gates for physical systems},
\newblock Phys. Rev. A \textbf{84}, 042315 (2011),
\newblock \doi{10.1103/PhysRevA.84.042315}.

\bibitem{SchaeferJCompP2017}
I.~Schaefer, H.~Tal-Ezer and R.~Kosloff,
\newblock \emph{Semi-global approach for propagation of the time-dependent
  {Schr{\"o}dinger} equation for time-dependent and nonlinear problems},
\newblock J. Comput. Phys. \textbf{343}, 368 (2017),
\newblock \doi{10.1016/j.jcp.2017.04.017}.

\bibitem{PalaoPRA13}
J.~P. Palao, D.~M. Reich and C.~P. Koch,
\newblock \emph{Steering the optimization pathway in the control landscape
  using constraints},
\newblock Phys. Rev. A \textbf{88}, 053409 (2013),
\newblock \doi{10.1103/PhysRevA.88.053409}.

\bibitem{ReichJMO14}
D.~M. Reich, J.~P. Palao and C.~P. Koch,
\newblock \emph{Optimal control under spectral constraints: Enforcing
  multi-photon absorption pathways},
\newblock J. Mod. Opt. \textbf{61}, 822 (2014),
\newblock \doi{10.1080/09500340.2013.844866}.

\bibitem{LapertPRA09}
M.~Lapert, R.~Tehini, G.~Turinici and D.~Sugny,
\newblock \emph{Monotonically convergent optimal control theory of quantum
  systems with spectral constraints on the control field},
\newblock Phys. Rev. A \textbf{79}, 063411 (2009),
\newblock \doi{10.1103/PhysRevA.79.063411}.

\bibitem{BukinPRX2018}
M.~Bukov, A.~G.~R. Day, D.~Sels, P.~Weinberg, A.~Polkovnikov and P.~Mehta,
\newblock \emph{Reinforcement learning in different phases of quantum control},
\newblock Phys. Rev. X \textbf{8}, 031086 (2018),
\newblock \doi{10.1103/PhysRevX.8.031086}.

\bibitem{BartanaJCP93}
A.~Bartana, R.~Kosloff and D.~J. Tannor,
\newblock \emph{Laser cooling of molecular internal degrees of freedom by a
  series of shaped pulses},
\newblock J. Chem. Phys. \textbf{99}, 196 (1993),
\newblock \doi{10.1063/1.465797}.

\bibitem{OhtsukiJCP99}
Y.~Ohtsuki, W.~Zhu and H.~Rabitz,
\newblock \emph{Monotonically convergent algorithm for quantum optimal control
  with dissipation},
\newblock J. Chem. Phys. \textbf{110}, 9825 (1999),
\newblock \doi{10.1063/1.478036}.

\bibitem{GoerzQST2018}
M.~H. Goerz and K.~Jacobs,
\newblock \emph{Efficient optimization of state preparation in quantum networks
  using quantum trajectories},
\newblock Quantum Sci. Technol. \textbf{3}, 045005 (2018),
\newblock \doi{10.1088/2058-9565/aace16}.

\end{thebibliography}

\end{document}